\documentclass[aps,prb,amsmath,amssymb,showpacs,reprint,onecolumn]{revtex4-2}
\usepackage{graphicx}
\usepackage{bm}
\usepackage{braket}
\usepackage[colorlinks=true,linkcolor=blue,citecolor=blue,urlcolor=blue]{hyperref}

\begin{document}

\title{Tilt-controlled Drude anisotropy in nodal-line semimetals: logarithmic enhancement, interband correction, gauge consistency, and the Landau-damping window}

\author{Wei Li}
\altaffiliation{Corresponding author: wliustc@aust.edu.cn}
\affiliation{Center of Fundamental Physics, Anhui University of Science and Technology, Huainan 232001, China}
\author{Jing-Rong Wang}
\altaffiliation{Corresponding author: wangjr@hmfl.ac.cn}
\affiliation{High Magnetic Field Laboratory of Anhui Province, Chinese Academy of Sciences, Hefei 230031, China}

\date{\today}

\begin{abstract}
    We study the anisotropic charge response of a three-dimensional tilted nodal-line semimetal in which the tilt enters the spectrum through an identity term and acts as the model control parameter. Using the Kubo formula with line-integral vertices, we show that the tilt-induced deformation of the low-energy phase space produces parametrically different longitudinal and transverse Drude responses: the longitudinal weight grows logarithmically as $D_{zz}=B\ln(1/\delta)$ with the analytically determined coefficient $B=g/(2\pi)^2$---within the window in which the Fermi-surface tail lies inside the ultraviolet cutoff, while at a fixed physical cutoff the weight saturates instead at a cutoff-set plateau---whereas the transverse weight follows a stronger algebraic scaling $D_\perp\sim\delta^{-1}$; the leading $\sqrt\delta$ corrections to both asymptotic forms are obtained analytically. The interband polarization retains the $q^2$ long-wavelength structure, with a coefficient that is infrared-integrable through Pauli blocking at finite doping yet carries a slow ultraviolet logarithm; it renormalizes the collective mode multiplicatively and sets the interband absorption threshold $\omega_c=2\mu/(1+|\eta_z|)$. A self-consistent solution of the plasmon dispersion identifies a tilt-dependent damping-free window bounded by the interband continuum. We predict that the type-II extension of this window, within the stated bandwidth (Brillouin-zone-like) cutoff regularization, contains a candidate low-frequency acoustic-like branch outside the particle--hole continua for $\eta_z\gtrsim1.2$ over a finite window in momentum transfer.
\end{abstract}

\maketitle

\section{Introduction}
\label{sec:intro}

Nodal-line semimetals (NLSMs) are three-dimensional materials whose conduction and valence bands cross along closed loops, or nodal lines, in momentum space~\cite{bian2016,schoop2016}; such crossings are protected by the combination of inversion and time-reversal symmetry, or by nonsymmorphic space-group symmetries~\cite{fang2015nonsymm}, and the resulting phases are reviewed in Refs.~\cite{burkov2016,armitage2018,fang2016cpb}. The drumhead surface states~\cite{schoop2016,xie2015} and the anisotropic bulk response distinguish NLSMs from point-node Weyl and Dirac semimetals; nodal-line materials have been realized and characterized in photoemission and transport, with first-principles guidance from Ca$_3$P$_2$~\cite{xie2015}, the antiperovskite Cu$_3$PdN~\cite{yu2015} and CaAgAs~\cite{wang2017caagw}, and the experimentally confirmed nodal-line family~\cite{schoop2016,bian2016}, while layered nonsymmorphic semimetals such as ZrSiS host nodal lines together with surface floating bands~\cite{topp2017}, with the two-dimensional Dirac bands producing a flat optical conductivity~\cite{schilling2017} and optical spectroscopy resolving the nodal-line signatures in $\mathrm{NbAs_2}$~\cite{shao2019optical}. A finite tilt of the nodal ring breaks particle--hole symmetry in the spectrum and, beyond a critical tilt, drives a Lifshitz-like transition from a type-I to a type-II regime in which the Fermi surface develops open pockets; the type-I/II tilt classification, developed for Weyl and Dirac semimetals~\cite{soluyanov2015,armitage2018} and completed by the type-III case~\cite{li2021typeiii}, is what makes the tilt a continuously tunable control knob. Single-particle transport and optical responses in tilted nodal-line and tilted Dirac systems have been studied~\cite{wareham2023,moors2019,wang2021ahal,ekstrom2021,tan2022,jalalimola2021}, with cold-atom shaken-lattice analogues~\cite{jotzu2014} and first-principles guidance~\cite{weng2016}; the collective response of untilted 3D NLSMs was established by Rhim and Kim~\cite{rhim2016}, with further studies of dispersive nodal-line and topological-semimetal collective excitations~\cite{losic2022,xue2023,islam2021}, of the electrodynamics on the Fermi cyclides~\cite{ahn2017} and the universality of plasmon excitations in three-dimensional Dirac semimetals~\cite{kharzeev2015}, of the nodal-line optical and thermoelectric response~\cite{barati2017,barati2020}, of plasmons in two-dimensional nonsymmorphic nodal lines~\cite{cao2023}, of the experimental plasmon response of $\mathrm{TiSe_2}$~\cite{lin2022}, and of correlated instability channels of quadratic and cubic nodal-line fermions~\cite{wang2020nlsp}; in our earlier work the RPA density scaling of plasmons in quadratic and cubic nodal-line semimetals was established~\cite{liwang2026}. Tilted Weyl and Dirac semimetals show rich plasmon physics, including a gapless undamped mode in the type-II regime~\cite{sadhukhan2020}, acoustic collective branches in the type-I regime~\cite{afanasiev2021}, 3D Dirac plasmons in the type-II semimetal $\mathrm{PtTe_2}$~\cite{politano2018}, topologically nontrivial interband plasmons in the type-II Weyl semimetal $\mathrm{MoTe_2}$~\cite{jia2020}, and hyperbolic plasmons in massive tilted two-dimensional Dirac materials~\cite{mojarro2022}; the 3D nodal-line case with tilt, however, has not, to our knowledge, been analyzed for its collective (plasmon/Landau-damping) response at finite $q$; the present work shares its model Hamiltonian with the single-particle anomalous-Hall optical study of Ref.~\cite{wang2021ahal} but addresses a different response channel.

Three specific gaps motivate this work. First, the longitudinal Drude weight of a tilted nodal-line semimetal grows logarithmically as $\delta\equiv1-\eta_z\to0^+$ but the coefficient $B$ was not fixed analytically and the transverse weight has not been analyzed. Second, in the polarizability of three-dimensional Dirac nodal-line systems a \emph{resonant} interband contribution with cubic wave-vector dependence has recently been identified in the long-wavelength limit, for chemical potentials approaching the band edge, in contrast to the quadratic ($q^2$) dependence of the standard intraband and interband processes~\cite{pandey2025}; the $q^2$ coefficient at the untilted point ($\eta_{\rm RK}=\pi/2$ in the Rhim--Kim geometry~\cite{rhim2016}, where the polarizability reduces to the graphene-like form) has not been extended to the tilted case, and the role of Pauli blocking at finite doping in setting the interband coefficient and absorption threshold has not been identified. Third, the gauge consistency of the response tensor with a hard cutoff requires line-integral vertices whose consequences for the Ward identities have not been worked out.

These gaps are addressed within a single framework. The key physical insight is that the tilt $\eta_z$ controls the Drude weights through the geometry of the Fermi-surface tail, but through \emph{distinct} mechanisms in the longitudinal and transverse channels: under the same tail measure $dk_z$, the longitudinal weight per unit $k_z$ scales as $1/|k_z|$ in the matching tail, so that the tail integrates to a logarithm, whereas the transverse weight per unit $k_z$ tends to a constant on the same tail, so that the tail contributes extensively and gives the power law. At finite wavevector the collective response additionally requires the interband polarization (IR-integrable by Pauli blocking at finite $\mu$, but carrying a slow ultraviolet logarithm) and density-channel gauge-consistent vertices, which are independent requirements rather than consequences of a single ``master parameter.'' The central results of this work are as follows. The longitudinal Drude weight grows as $D_{zz}=B\ln(1/\delta)$ with the analytically determined coefficient $B=g/(2\pi)^2$, with the same framework closing the subleading $\sqrt\delta$ correction analytically. The transverse weight follows a distinct power law $D_\perp\sim\delta^{-1}$, driven by the unprotected transverse velocity on the tail; the power law holds in the tail-inside-cutoff window, and the weight saturates at fixed cutoffs beyond it. The interband polarization scales as $P_2\propto q^2$---as at the untilted point~\cite{rhim2016}---with a UV-logarithmic coefficient $C^{\rm inter}$ whose $\eta_z$ independence---a blindness of the geometric weight, not a tilt dependence---and multiplicative RPA role we establish here; it corrects the RPA plasmon frequency multiplicatively and subleadingly, $\omega_p^2=D_{zz}/(1-C^{\rm inter})$, and the plasmon lies below the interband absorption threshold $\omega_c=2\mu/(1+|\eta_z|)$. The same kinematics fixes the Landau-damping window: for $\mathbf q\parallel\hat z$ the intraband continuum occupies $[0,(1+\eta_z)q]$, and the long-wavelength estimate $q^*=\omega_p^{(0)}/(1+\eta_z)$ would place the mode above that continuum only for $q<q^*$; a self-consistent evaluation at finite $q$ shows, however, that the mode rises with $q$ and stays above the intraband continuum, so the intraband channel does not open and the damping-free window is bounded instead by the interband edge, at a tilt-dependent wavevector $q_{\rm inter,self}(\eta_z)$ that decreases as the tilt approaches unity (Table~\ref{tab:qinter}). In the type-II regime ($\eta_z>1$) the intraband continuum acquires a low-frequency gap, and a low-frequency acoustic-like branch enters the resulting low-gap window for $\eta_z\gtrsim1.2$ within the stated bandwidth (Brillouin-zone-like) cutoff regularization.

The density-channel Ward identity is analytically exact within the continuum model at momentum transfer below the boundary-crescent validity bound (Appendix~\ref{sec:ward-appendix}) and finite frequency within the hard-cutoff domain that contains the Fermi surface; the transverse channel retains a scheme-dependent residual, so gauge consistency is claimed only for the density-longitudinal channel. Type-II results are quoted within the stated Brillouin-zone regularization, which provides the finite-size platform for the open Fermi-surface pockets.

The paper is organized as follows. Section~\ref{sec:model} introduces the model and the Fermi-surface tail geometry, and Sec.~\ref{sec:drudebench} reduces the Kubo formula to one-dimensional Drude integrals and fixes the conventions. Section~\ref{sec:logenh} establishes the longitudinal logarithmic enhancement $D_{zz}=B\ln(1/\delta)$; Sec.~\ref{sec:transverse} derives the transverse power law and its crossover function. Section~\ref{sec:interband} treats the interband polarization $P_2\propto q^2$, its ultraviolet structure, and the absorption threshold; Sec.~\ref{sec:plasmons} discusses the collective plasmon response, its absorption threshold and its Landau-damping window (Sec.~\ref{sec:damping-window}). Section~\ref{sec:ward} establishes the density-channel Ward identity (with the detailed derivation in the appendices). The technical derivations are collected in the appendices, and Sec.~\ref{sec:conclusion} summarizes the results and limitations.

Unless stated otherwise, we adopt dimensionless units $\hbar=v_0=v_z=k_0=1$, energy in units of $\mu$, and the physical spin degeneracy $g=2$ for all reported physical quantities (Drude weights, interband polarization, dielectric function, and plasmon frequency); the response-tensor components of Sec.~\ref{sec:ward} are computed with $g=1$ as a bookkeeping convention (their residuals and ratios are $g$-independent). We also set the Coulomb coupling $4\pi e^2/\epsilon_b=1$ in these units, so the RPA plasmon frequency is $\omega_p^2=D_{zz}/(1-C^{\rm inter})$ [restoring the coupling $V=4\pi e^2/\epsilon_b$ gives $\omega_p^2=V D_{zz}/(1-V\,C^{\rm inter})$: the coupling multiplies both the Drude numerator and the interband term in the denominator]. The factor-of-two relation between the $g=1$ and $g=2$ conventions is fixed in Sec.~\ref{sec:drudebench}; the Coulomb coupling $g_c$ of Sec.~\ref{sec:plasmons} is a distinct quantity, not to be confused with the degeneracy $g$.

\section{Model and Fermi-surface geometry}
\label{sec:model}

\subsection{Hamiltonian and band structure}

We consider the parallel-tilted nodal-ring (PTNR) model---the parallel-tilted form of the nodal-ring model of Ref.~\cite{burkov2011}---used for the 3D tilted case in Ref.~\cite{wang2021ahal} and in Eq.~(11) of Ref.~\cite{mandal2026},
\begin{equation}
    H(\mathbf k) = \mathbf d(\mathbf k)\cdot\bm\sigma + \eta_z k_z\,I,
    \qquad
    \mathbf d(\mathbf k) = \bigl(k_\perp-k_0,\;-k_z,\;0\bigr),
    \label{eq:H}
\end{equation}
with $k_\perp=\sqrt{k_x^2+k_y^2}$ and the tilt vector along $z$. The tilt term is proportional to the identity and therefore shifts the two bands rigidly without mixing them, in contrast to models in which the tilt enters through a $\sigma_z$ term---the type-II Weyl/Dirac $\sigma_z$-tilt variants~\cite{soluyanov2015,armitage2018}---where the tilt opens a Fermi surface at the node. Diagonalizing Eq.~\eqref{eq:H} gives
\begin{equation}
    \varepsilon_{\mathbf k,s} = \eta_z k_z + s\,E_{\mathbf k},
    \qquad
    E_{\mathbf k} = \sqrt{(k_\perp-k_0)^2+k_z^2},
    \qquad s=\pm 1.
    \label{eq:spec}
\end{equation}
The nodal ring lies at $k_\perp=k_0$, $k_z=0$. Throughout we set $\hbar=v_0=v_z=k_0=1$ and measure energy in units of $\mu$. The conventions for the spin degeneracy $g$---the physical $g=2$ for all reported quantities and the $g=1$ bookkeeping for the response-tensor components of Sec.~\ref{sec:ward}---are fixed once in Sec.~\ref{sec:drudebench} and used without restatement.

\subsection{Fermi surface in the weak-tilt limit}

The Fermi surface is determined by $\varepsilon_{\mathbf k,s}=\mu$. With $\mu=1$ (the ring scale $k_0$), the occupied states satisfy $\varepsilon_{\mathbf k,-}\le \mu\le \varepsilon_{\mathbf k,+}$, and the Fermi-surface integration domain in the $(k_z,k_\perp)$ plane is, for type-I tilt ($\delta>0$),
\begin{equation}
    \mathcal I = \bigl[-\mu/\delta,\;\mu/(1+\eta_z)\bigr],
    \qquad
    \delta\equiv 1-\eta_z,
    \label{eq:domain}
\end{equation}
Throughout, the scalar $\delta$ denotes the tilt parameter $1-\eta_z$ (which is negative in the type-II regime, where expressions such as $\ln(1/\delta)$ are not defined and the type-II analysis of Sec.~\ref{sec:damping-window} is used instead); the Dirac distribution appearing in the Kubo formulas is always written with an explicit argument, $\delta(\cdot)$, and never without one. For type-II tilt ($\delta<0$) the Fermi surface is open and the domain is $\mathcal I_{\rm II}=(-\infty,\ \mu/(1+\eta_z)]$ (the radial-discriminant condition is satisfied for all $k_z<0$), as discussed in Sec.~\ref{sec:damping-window}; Eq.~\eqref{eq:domain} is used for the type-I analysis of Secs.~\ref{sec:logenh}--\ref{sec:transverse}. Two features of Eq.~\eqref{eq:domain} dominate the weak-tilt physics studied in Secs.~\ref{sec:logenh}--\ref{sec:transverse}. First, the interval is asymmetric: the long tail extends to $k_z=-\mu/\delta$ on the \emph{negative} side, diverging as $\delta\to 0^+$, while the positive endpoint remains finite at $\mu/(1+\eta_z)\to \mu/2$. Second, within $\mathcal I$ the radial root
\begin{equation}
    k_\perp^{(\pm)}(k_z) = 1 \pm \sqrt{(\mu-\eta_z k_z)^2-k_z^2}
    \label{eq:kperp}
\end{equation}
give the two radial roots of the Fermi-surface equation. Only the $s=+$ band crosses $\mu$ for the type-I parameters used here ($\mu=k_0$; the $s=-$ equation has no solution in $k_\perp\ge0$), but that band has \emph{two} radial roots, both summed in the Drude integrals below through the weight $\mathcal W=\sum_{\sigma:k_{\perp,\sigma}\ge0}k_{\perp,\sigma}/R$. The long tail is the source of the logarithmic enhancement of the longitudinal Drude weight (Sec.~\ref{sec:logenh}) and, through the transverse velocity $v_x\propto (k_\perp-1)/E_{\mathbf k}\to 0$ on the tail (the $E^{-1}$ factor decays, while $(k_\perp-1)/k_\perp\to 1$), of the distinct $\delta^{-1}$ scaling of the transverse weight (Sec.~\ref{sec:transverse}). The interval $\mathcal I$ is a property of the continuous low-energy model: there is no lattice and hence no Brillouin zone in the analytic calculation, and the endpoints are fixed by the Fermi-surface equation alone. This distinction matters only where a cutoff is introduced. Within the window where the Fermi-surface tail lies inside the cutoff, the logarithmic law of Sec.~\ref{sec:logenh} has a cutoff-independent leading coefficient $B$, the cutoff entering only the $O(1)$ constant; at a fixed physical cutoff $\Lambda$ (the Brillouin-zone boundary in a lattice) the tail is truncated at $y_{\max}=\delta\Lambda$ in the tail coordinate of Sec.~\ref{sec:model}, and $D_{zz}$ saturates at a plateau $B\ln(\Lambda/c)+O(1)$ independent of $\delta$---the saturation regime of a lattice model with a fixed Brillouin-zone boundary. The type-I results require the Fermi-surface tail to lie inside the cutoff, and the hard-cutoff (bandwidth) platform of Sec.~\ref{sec:damping-window} is the scheme adopted for type-II.

\begin{figure}[t]
    \centering
    \includegraphics[width=0.98\columnwidth]{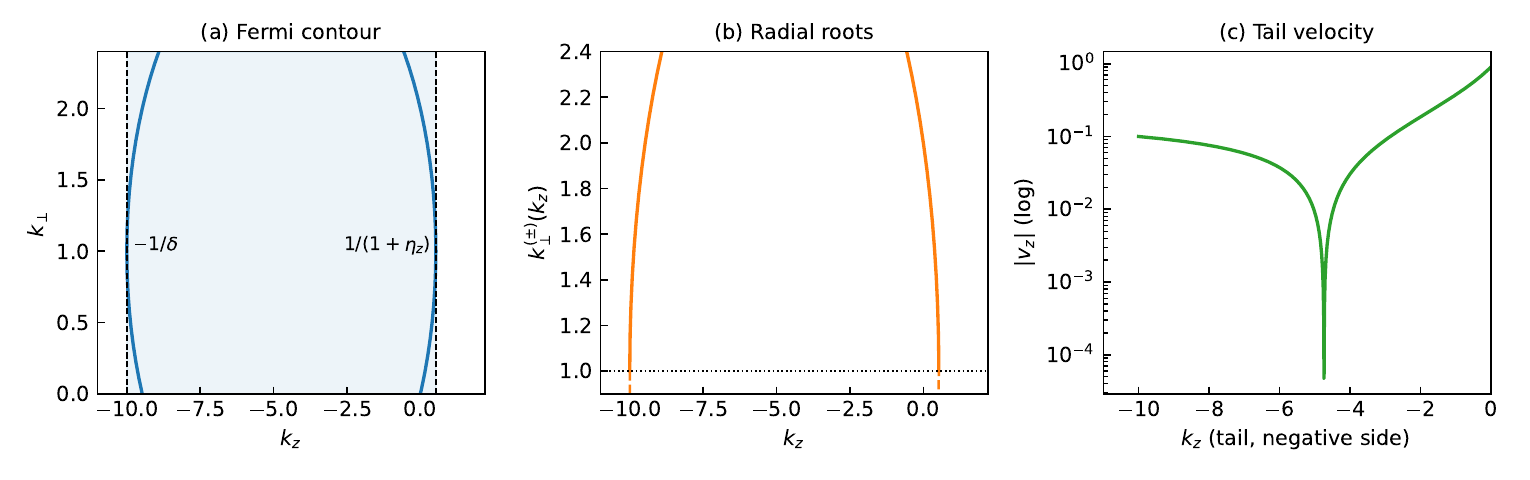}
    \caption{Fermi-surface geometry of the tilted nodal ring at $\mu=1$. (a) Fermi contour in the $(k_z,k_\perp)$ plane for $\eta_z=0.9$, showing the asymmetric integration interval $\mathcal I=[-\mu/\delta,\mu/(1+\eta_z)]$ (shaded) with the long tail on the negative-$k_z$ side. (b) Radial roots $k_\perp^{(\pm)}(k_z)$ (solid: outer $+$; dashed: inner $-$), symmetric about the dotted line $k_\perp=1$, the nodal-ring radius. (c) Longitudinal velocity magnitude $|v_z|=|\partial_{k_z}\varepsilon_{{\mathbf k},+}|$ of the occupied ($s=+$) band along the tail, on a logarithmic scale: it decays from $O(1)$ near the origin, vanishes at $k_z=\mu/(\eta_z-1/\eta_z)\approx-4.74$ (for $\eta_z=0.9$, the zero of $v_z$), and reaches $|\!-\delta|=0.1$ at the tail end $k_z=-\mu/\delta$ (the dashed line marks $v_z=0$).}
    \label{fig:fs}
\end{figure}

Figure~\ref{fig:fs} summarizes the Fermi-surface geometry: the tilted ring in the $(k_z,k_\perp)$ plane, the interval $\mathcal I$ with its asymmetric endpoints, and the boundary-layer structure near the long-tail end that controls the endpoint contributions to the Drude weight.

The decomposition of the Drude weight into the bulk, matching-tail, and endpoint contributions follows the matched-asymptotics scheme of Sec.~\ref{sec:logenh}; the constants that control the endpoint layers are computed analytically in Sec.~\ref{sec:logenh}.

\subsection{Tail coordinate and boundary-layer structure}

For the asymptotic analysis it is convenient to parametrize the long tail by $y\equiv-\delta k_z$, so that the negative segment $k_z\in[-\mu/\delta,0]$ maps (order-reversingly, $dk_z=-\delta^{-1}dy$) to the compact interval $y\in[0,\mu]$, while the positive side $k_z\in[0,\mu/(1+\eta_z)]$ is a fixed-width segment. In these coordinates the radial root and the velocity become
\begin{equation}
    E = \frac{y+\delta(1-y)}{\delta},
    \qquad
    R^2 = \frac{(1-y)[2y+\delta(1-y)]}{\delta},
    \qquad
    v_z = \frac{\delta[(1-\delta)(1-y)-y]}{y+\delta(1-y)},
    \label{eq:ytail}
\end{equation}
with the measure $dk_z=\delta^{-1}dy$. Two structural features follow directly. First, the tail integrand factorizes into a $\delta$-independent leading part $(1-2y)^2/y$ plus a $O(\sqrt\delta)$ correction that carries the slow-convergence contamination of the log slope (Sec.~\ref{sec:logenh}). Second, the two endpoints of the tail scale differently: the tail start at $y=0$ (equivalently $k_z=0$) supports a boundary layer of width $O(\delta)$ whose integrated weight is $O(1)$ and is controlled by the analytic constant $L_-$, while the far end at $y=1$ ($k_z=-\mu/\delta$; recall $y\equiv-\delta k_z$ and $\mu=1$) is a boundary layer of width $O(\delta)$ and weight $O(\delta)$, controlled by $\delta A_+$ (Sec.~\ref{sec:logenh}). This three-scale structure---bulk, logarithmic tail, and endpoint layers---is realized as the four-region decomposition of the integration domain used in Sec.~\ref{sec:logenh} (fixed interior, matching tail, far-end bulk, and the two endpoint layers) and is the organizing principle of Secs.~\ref{sec:logenh}--\ref{sec:transverse}.

\section{Kubo reduction and conventions}
\label{sec:drudebench}

\subsection{Kubo formula and reduced one-dimensional integral}

The longitudinal Drude weight is defined by the zero-frequency limit of the Kubo formula~\cite{yan2016,sadhukhan2020} as the positive quantity
\begin{equation}
    D_{zz} = g\sum_{s=\pm}\int\frac{d^3k}{(2\pi)^3}\,v_{z,\mathbf k,s}^2\,\delta(\mu-\varepsilon_{\mathbf k,s}),
    \label{eq:Dzz}
\end{equation}
where the sum runs over both bands and the delta function selects whichever band crosses $\mu$; for the weak-tilt type-I parameters of this paper ($\mu=k_0$) the Fermi surface lies entirely on the $s=+$ (conduction) band, and the $s=-$ equation $\varepsilon_{\mathbf k,-}=\mu$ has no solution in $k_\perp\ge0$. (The response tensor of Sec.~\ref{sec:ward} uses the opposite sign convention to Eq.~\eqref{eq:Dzz}, $\Pi^{00}(0,0)=+\partial n/\partial\mu$; the two are distinct objects and the signs are stated separately.)
The band velocity is
\begin{equation}
    v_{z,\mathbf k,s} = \frac{\partial\varepsilon_{\mathbf k,s}}{\partial k_z}
    = \eta_z + s\,\frac{k_z}{E_{\mathbf k}},
    \label{eq:vz}
\end{equation}
where the minus sign is the Kubo convention and the physical weight is positive. For the tilted ring the delta function restricts the integral to the single Fermi sheet $k_\perp^{(+)}(k_z)$ of Eq.~\eqref{eq:kperp}, and the azimuthal integral contributes $2\pi k_\perp$, reducing Eq.~\eqref{eq:Dzz} to the analytically exact one-dimensional form
\begin{equation}
    D_{zz}(\eta_z) = \frac{g}{(2\pi)^2}\int_{\mathcal I} dk_z\,
    \mathcal W(k_z)\,E\,v_{z,\mathbf k,+}^2,
    \qquad
    \mathcal W = \sum_{\sigma:\,k_{\perp,\sigma}\ge0}\frac{k_{\perp,\sigma}}{R},
    \label{eq:Dzz_reduced}
\end{equation}
with the integration domain $\mathcal I$ of Eq.~\eqref{eq:domain} and $|\nabla_{\mathbf k}\varepsilon|=\sqrt{(\partial\varepsilon/\partial k_\perp)^2+(\partial\varepsilon/\partial k_z)^2}$. The step-by-step reduction (delta-function evaluation, azimuthal integration, domain specification) is given in Appendix~\ref{app:drude}; Eq.~\eqref{eq:Dzz_reduced} is the exact starting point of the matched-asymptotics analysis of Sec.~\ref{sec:logenh}.

The reduced one-dimensional form closes at the untilted limit, where Eq.~\eqref{eq:Dzz_reduced} evaluates to $D_{zz}(0)=g/(4\pi)$ and the static density response to $\partial n/\partial\mu=g/(2\pi)$, and it fixes the relation between the two degeneracy conventions used in this paper---the physical $g=2$ for all reported quantities and the $g=1$ bookkeeping for the response-tensor components of Sec.~\ref{sec:ward}---which differ exactly by the factor $g$.

The full derivation of Eq.~\eqref{eq:Dzz_reduced} is given in Appendix~\ref{app:drude}.

The reduced form of Eq.~\eqref{eq:Dzz_reduced} is the analytically exact statement of the Drude weight for the tilted ring: no further approximation enters between the Kubo formula and the one-dimensional integral. Two consequences follow. First, the integral is the natural object for analytic control, because the velocity, the radial root, and the domain are all elementary functions of $k_z$; the matched-asymptotics expansion of Sec.~\ref{sec:logenh} operates directly on this form. Second, the reduced form makes explicit the role of the spin degeneracy $g$: it enters as an overall factor, so all ratios of Drude weights (e.g.\ the longitudinal-to-transverse anisotropy) are $g$ independent, while absolute values require the stated convention; the functional form $g/(2\pi)^2$ does not depend on computational conventions, while its numerical value scales linearly with the physical degeneracy $g$.

\section{Logarithmic enhancement of the longitudinal Drude weight}
\label{sec:logenh}

\subsection{Four-region decomposition}

The exact reduced integral, Eq.~\eqref{eq:Dzz_reduced}, is decomposed into four regions that realize the three scales of Sec.~\ref{sec:model}: a fixed interior of order $\mu^2$ and a far-end bulk of order $\mu^2$ (the two parts of the bulk scale), a matching tail that carries the logarithm (the logarithmic tail), and the endpoint layers at the two ends of the tail, whose contributions are $O(\mu^2)$ at the tail start ($L_-$, finite after the $1/\sqrt\delta$ divergence is absorbed into the leading log) and $O(\delta\,\mu^2)$ at the far end ($\delta A_+$). Introducing the tail coordinate $y\equiv-\delta k_z$ of Sec.~\ref{sec:model} (the long tail $k_z\in[-1/\delta,0]$ maps to $y\in[0,1]$ with $k_z=-1/\delta\to y=1$ and $k_z=0\to y=0$, and $|dk_z/dy|=1/\delta$ flips the integration direction), the tail integrand of the Drude formula becomes, at leading order,
\begin{equation}
    \frac{1}{\delta}\,\mathcal W E\, v_z^2
    = \frac{(1-2y)^2}{y}
    + O\!\Bigl(\frac{\sqrt\delta}{y^{3/2}\sqrt{1-y}} + \frac{\delta}{y^2}\Bigr),
    \label{eq:eqn11}
\end{equation}
where $\mathcal W=\sum_\sigma k_{\perp,\sigma}/R$ is the delta-function weight and $R^2=(1-\eta_z k_z)^2-k_z^2$ is the radial discriminant. Integrating Eq.~\eqref{eq:eqn11} over $y\in[\delta,1]$ gives the main result of this section:
\begin{equation}
    D_{zz} = B\ln\frac{1}{\delta} + O(1),
    \qquad
    B = \frac{g}{(2\pi)^2} = \frac{g}{4\pi^2}.
    \label{eq:eqn13}
\end{equation}
The coefficient is fixed purely by the tail geometry: the factor $g/(2\pi)^2$ combines the spin degeneracy $g$ and the phase-space measure $1/(2\pi)^2$ of the $(y,k_\perp)$ tail coordinates; no other dimensionless model parameter enters $B$ at the reference point $\mu=k_0=1$ used throughout (in physical units the logarithmic coefficient carries the overall scale $(\mu/k_0)^2$---the untilted weight itself scales as $g k_0\mu/4\pi$, i.e.\ linearly in $\mu$---so the quoted $B$ is the coefficient at $\mu=k_0$). The $O(1)$ term of Eq.~\eqref{eq:eqn13} is not a single closed-form constant; its composition and numerical behavior are described in Appendix~\ref{app:drude}.

The endpoint layers contribute constants that are $\delta$ independent (at the tail start) or of order $\delta$ (at the far end):
\begin{equation}
    L_-(c) = \ln(1+c) + 2\arctan\sqrt{1+2c} - \frac{\pi}{2},
    \qquad
    A_+(c) = c + \tfrac12 + \sqrt{2c}\quad(c\ge\tfrac12),
    \label{eq:endpoint}
\end{equation}
where $c$ is the integration cap of the layer coordinates (the rescaled distances $s=y/\delta$ and $w=(1-y)/\delta$ from the two endpoints), taken at the matching cap $c=1$; the layer contributions are $(g/(2\pi)^2)L_-(c)$ from the tail-start layer (the $y=0$ end) and $(g/(2\pi)^2)\,\delta A_+(c)$ from the far-end layer (the $y=1$ end). The endpoint-layer amplitudes are the analytic constants of Eq.~\eqref{eq:endpoint}.

The physical origin of the logarithm is the divergence of the weight per unit $k_z$ at the tail start: in the tail coordinate the leading integrand of Eq.~\eqref{eq:eqn11} is $(1-2y)^2/y$, i.e.\ it scales as $1/|k_z|$ in the matching tail---the energy factor grows as $E\sim|k_z|$ while $v_z^2$ decays as $y^{-2}$, so that the combination $\delta^{-1}\mathcal W E v_z^2$ carries a $1/y$ singularity---and integrating it over the tail gives a logarithm rather than a power. This is the three-dimensional analogue of the geometry-driven response enhancement identified for two-dimensional nodal rings by Rahimpoor and Abedinpour~\cite{rahimpoor2024}, but with a crucial difference: in 2D the Drude weight is independent of the chemical potential because the ring geometry fixes the phase space, whereas here the enhancement is tied to the divergent tail and is therefore controlled by the tilt alone. The coefficient $B=g/(2\pi)^2$ contains no Fermi-surface curvature data, which is the signature of a purely kinematic (tail-measure) mechanism. The logarithmic law applies in the type-I regime below the transition; at the transition itself the tail contribution reorganizes into the open-pocket Fermi surface of the type-II regime (Sec.~\ref{sec:damping-window}). (The logarithmic law is a window statement: it requires the tail to lie inside the UV cutoff, $\delta\gg1/\Lambda$, and the tail to fit inside the BZ, $\delta\gtrsim k_0/\Lambda_{\rm BZ}$; see Sec.~\ref{sec:conclusion} for the accessible window.)

\subsection{Subleading $\sqrt\delta$ correction law}

The next-to-leading correction to Eq.~\eqref{eq:eqn13} is of order $\sqrt\delta$ and comes from the upper edge of the tail integral of the $\sqrt\delta$ term in Eq.~\eqref{eq:eqn11}. The effective log slope $B_{\rm eff}(\delta)\equiv dD_{zz}/d\ln(1/\delta)$ therefore approaches $B$ as
\begin{equation}
    B_{\rm eff}(\delta) - B = \tilde C_{1/2}\sqrt\delta + o(\sqrt\delta),
    \label{eq:eqn20}
\end{equation}
and the prefactor is obtained in closed analytic form by the $\theta=\arcsin\sqrt y$ substitution applied to the leading tail integrand $(1-2y)^2\sqrt\delta/(\sqrt2\,y^{3/2}\sqrt{1-y})$; with the exact antiderivative $F(y)=-\sqrt2[\cot\theta+2\theta+\sin2\theta]$, the lower-endpoint contribution is $\delta$ independent and the upper edge $y_1$ leaves
\begin{equation}
    \tilde C_{1/2} = -B\sqrt2\,\bigl[\cot\theta_1 + 2\theta_1 + \sin 2\theta_1\bigr],
    \qquad
    \theta_1=\arcsin\sqrt{y_1^{(0)}}=\arcsin\sqrt{[1-\sqrt{1-4u}]/2},
    \label{eq:eqn22}
\end{equation}
where $u=\tfrac{\delta}{2}(\Lambda_\perp-1)^2$ and $x=\delta\Lambda_z$ are the scaled transverse and longitudinal cutoffs and $y_1^{(0)}=[1-\sqrt{1-4u}]/2$ is the small root of $y(1-y)=u$ (the fixed-$u$ matched-asymptotic form requires $y_1^{(0)}<1$, i.e.\ $u<1/4$, which at the reference transverse cutoff $\Lambda_\perp=4$ confines the \emph{fixed-$u$ matched-asymptotic} analysis to $\delta<0.056$; this is a mathematical validity bound of the expansion, to be distinguished from the physical-cutoff window $\delta\gtrsim k_0/\Lambda$ of Sec.~\ref{sec:model}). The fixed-$u$ form holds the \emph{scaled} cutoffs fixed, so the physical cutoffs $\Lambda_z=x/\delta$ and $\Lambda_\perp=1+\sqrt{2u/\delta}$ grow as $\delta\to0$; it is the family in which the logarithmic law of Sec.~\ref{sec:logenh} is defined, not fixed physical cutoffs (at a fixed physical cutoff $D_{zz}$ saturates; Sec.~\ref{sec:model}). The $\sqrt\delta$ term controls the leading pre-asymptotic correction to the logarithmic behavior, and $B_{\rm eff}$ approaches $B$ only after this correction is removed.

\section{Transverse Drude weight}
\label{sec:transverse}

The transverse Drude weight $D_\perp=2D_{xx}$ is obtained from the reduced formula,
\begin{equation}
    D_{xx} = \frac{g}{2(2\pi)^2}\int_{\mathcal I} dk_z\,\mathcal W_x,
    \qquad
    \mathcal W_x = \sum_{\sigma} \frac{k_{\perp,\sigma}}{R}\,E^{-1}\,\bigl[(k_\perp-1)^2\bigr],
    \label{eq:Dxx}
\end{equation}
whose tail asymptotics in the $y$ coordinate gives the leading scaling
\begin{equation}
    D_{xx}^{(-)} = \frac{g}{2(2\pi)^2}\,\frac{1}{\delta}
    \Bigl[1 + \frac{\pi}{2}\sqrt{2\delta} + \cdots\Bigr],
    \qquad
    D_\perp^{(-)} \sim \frac{g}{(2\pi)^2}\,\frac{1}{\delta},
    \label{eq:CLM-B-01}
\end{equation}
establishing the $\delta^{-1}$ scaling of the transverse weight in the weak-tilt limit, unlike the logarithmic enhancement of $D_{zz}$ (Sec.~\ref{sec:logenh}). The mechanism is the saturation of the transverse tail weight: on the long tail the radial factor $k_\perp-1=R$ grows as $\delta^{-1/2}$ while $E_{\mathbf k}$ grows as $\delta^{-1}$, so the integrand of Eq.~\eqref{eq:Dxx}, $\mathcal W_x=E^{-1}(1+R)R$, tends to a \emph{constant} across the tail instead of the $1/|k_z|$ singularity of the longitudinal channel; the extensive tail measure then yields $\delta^{-1}$ rather than a logarithm.

\begin{figure}[t]
    \centering
    \includegraphics[width=0.85\columnwidth]{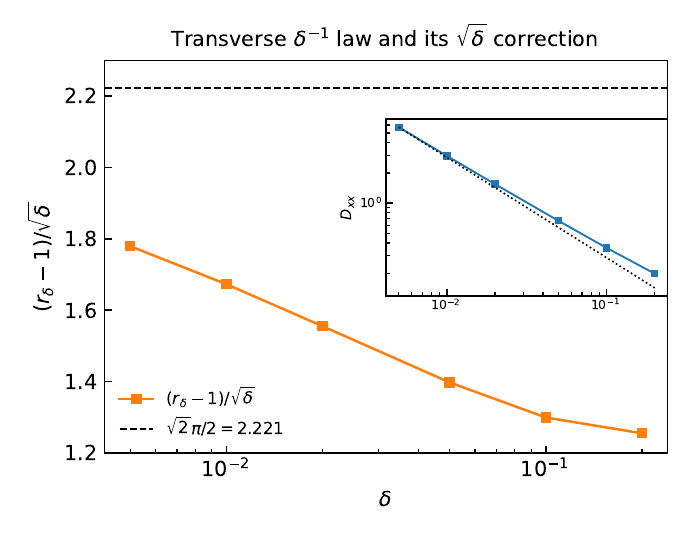}
    \caption{Transverse scaling: the ratio $(r_\delta-1)/\sqrt{\delta}$, with $r_\delta=D_{xx}\delta/\mathcal T$ and $\mathcal T=g/[2(2\pi)^2]$, versus $\delta$; it rises monotonically toward the analytic $\sqrt2\pi/2=2.221$ (dashed), confirming the $\delta^{-1}$ power law and its analytic $\sqrt\delta$ correction (equivalently $r_\delta\to1$ as $\delta\to0$). Inset: the raw $D_{xx}(\delta)$ on log-log axes with a $\delta^{-1}$ slope guide.}
    \label{fig:dperp}
\end{figure}

The coefficient of the $\sqrt\delta$ correction in Eq.~\eqref{eq:CLM-B-01} is the analytic constant $\sqrt2\pi/2=2.221$; the transverse scaling approaches this value as $\delta\to0$ within the $O(\delta\ln(1/\delta))$ family (Fig.~\ref{fig:dperp}). For the double-cutoff problem, the tail contribution saturates at
\begin{equation}
    D_{xx}^{\rm tail} \sim 2\mathcal T\,\min\Bigl(\Lambda_z,\ \frac{(\Lambda_\perp-1)^2}{2}\Bigr), \qquad \mathcal T \equiv \frac{g}{2(2\pi)^2},
    \label{eq:CLM-B-02}
\end{equation}
This is the leading plateau of the thin-ring tail asymptotic; the $O(1)$ coefficient follows from the $\mathcal W_x$ profile across the tail and is given in closed form in Appendix~\ref{app:drude} [Eq.~\eqref{eq:app_sat}], where it evaluates to $2.37\mathcal T\min(\cdots)$ for the cutoffs used here. On the type-I side, then, as $\delta\to0^+$ at fixed cutoffs, the transverse weight saturates at a cutoff-bound plateau rather than diverging (the type-II regime $\eta_z>1$ corresponds to $\delta<0$ and is outside the weak-tilt expansion); the leading crossover function $F(x,u)$ and its relation to the exact cutoff integral $F_\delta(x,u)$ are derived in Appendix~\ref{app:drude}.

\section{Interband polarization and the plasmon gap}
\label{sec:interband}

The collective response of the tilted NLSM is not determined by the intraband Drude weights alone: at finite wavevector the interband polarization contributes a correction to the plasmon frequency. This section derives the $q^2$ scaling of the interband polarization in the 3D nodal-line geometry, establishes that its coefficient is infrared-integrable through Pauli blocking at finite doping while carrying a slow ultraviolet logarithm [Eq.~\eqref{eq:inter-uvlog}], and quantifies the (subleading) correction to $\omega_p$. Throughout we use $\mu>0$ (electron doping); the intrinsic $\mu\to0$ limit is discussed at the end of the section.

\subsection{Interband Lindhard response and Pauli blocking}

The interband part of the density response at temperature $T=0$ receives contributions from the two channels $P_{-+}$ (valence $\to$ conduction, the absorption channel at positive frequency) and $P_{+-}$ (conduction $\to$ valence). The dynamic density response and the current--current susceptibility of three-dimensional Dirac and Weyl semimetals have been mapped within the same random-phase-approximation vertex set~\cite{panfilov2014,thakur2018}; the nodal-ring geometry considered here builds on that machinery. In the tilted model with $\varepsilon_{\mathbf k,s}=\eta_z k_z+sE_{\mathbf k}$, the valence band satisfies $\varepsilon_{\mathbf k,-}=\eta_z k_z-E_{\mathbf k}\le-(1-|\eta_z|)E_{\mathbf k}<0<\mu$ for type-I tilt $|\eta_z|<1$, so it is always occupied; the conduction band is occupied only where $\eta_z k_z+E_{\mathbf k}<\mu$. The full long-wavelength interband polarization at momentum transfer $\mathbf q=(0,0,q_z)$ is therefore
\begin{equation}
    P_2(\mathbf q,\omega) = \frac{g}{(2\pi)^3}\int d^3k\,
    F_{+-}(\mathbf k,\mathbf k+\mathbf q)\,
    \Theta(\eta_z k_z+E_{\mathbf k}-\mu)\,
    \biggl[\frac{1}{\omega-2E_{\mathbf k}+i0^+}
        -\frac{1}{\omega+2E_{\mathbf k}+i0^+}\biggr],
    \label{eq:P2def}
\end{equation}
where the first bracket is the absorption channel $P_{-+}$ and the second the reverse channel $P_{+-}$ (real for $\omega>0$, with no absorption pole below threshold); the Pauli factor $\Theta(\eta_z k_z+E_{\mathbf k}-\mu)\equiv f_- - f_+$ (the valence band fully occupied and the conduction band occupied for $\varepsilon_{+,{\mathbf k}}<\mu$) restricts the transitions to final conduction states above the Fermi level, in the standard Lindhard combination $f_s-f_{s'}$. The coherence factor is $F_{+-}(\mathbf k,\mathbf k+\mathbf q)=|\langle u_{+,\mathbf k}|u_{-,\mathbf k+\mathbf q}\rangle|^2$. For $\mathbf q\parallel\hat z$ and $q_z\ll E_{\mathbf k}$, expanding the eigenvector overlap to second order gives
\begin{equation}
    F_{+-} = \frac{(k_\perp-1)^2 q_z^2}{4E_{\mathbf k}^4} + O(q_z^4),
    \label{eq:Fpm}
\end{equation}
where the linear term vanishes by reflection symmetry in $k_z$ (the $y$ component of $\mathbf d$ is odd under $k_z\to-k_z$, so the in-plane overlap acquires no linear term). Equation \eqref{eq:P2def} keeps both the occupation factor and the interband energy denominator, the latter written in the $\mathbf q\to0$ form $2E_{\mathbf k}$ (the coherence factor of Eq.~\eqref{eq:Fpm} retains the leading $q^2$ dependence); earlier Fermi-surface-reduced treatments that drop either ingredient are incomplete (Sec.~\ref{sec:interband-note}).

\subsection{Pauli blocking, infrared integrability, and the ultraviolet scale of the interband coefficient}

Expanding Eq.~\eqref{eq:P2def} at small $q_z$ gives $P_2=C^{\rm inter}(\omega,\eta_z)\,q_z^2+O(q_z^4)$ with
\begin{equation}
    C^{\rm inter}(\omega,\eta_z)
    = \frac{g}{4\pi^2}\int k_\perp dk_\perp dk_z\,
    \frac{(k_\perp-1)^2}{4E_{\mathbf k}^4}\,
    \Theta(\eta_z k_z+E_{\mathbf k}-\mu)\,
    \frac{4E_{\mathbf k}}{\omega^2-4E_{\mathbf k}^2},
    \label{eq:P2scaling}
\end{equation}
after the azimuthal integral. The Pauli factor controls the \emph{infrared} end of the integral: it excludes the near-ring states with $\eta_z k_z+E_{\mathbf k}<\mu$ (and since $\eta_z k_z\le|\eta_z|E_{\mathbf k}$, only states with $(1+|\eta_z|)E_{\mathbf k}>\mu$ contribute). Without it the near-ring region would contribute a boundary term, but in the 3D ring measure this region is integrable ($k_\perp dk_\perp dk_z\sim E\,dE\,d\theta$ with a finite small-$E$ integrand), so there is no infrared logarithm, at any $\mu\ge0$. The coefficient is nevertheless \emph{not} finite in the continuum model: at large momentum the Pauli factor tends to unity, and the physical half-plane $k_\perp\ge0$ ($u\equiv k_\perp-1\ge-1$) breaks the $\cos^3\theta$ symmetry of the angular weight, leaving $\int(1+E\cos\theta)\cos^2\theta\,d\theta=\pi/2+(4/3)E$ and hence a logarithmically ultraviolet-divergent coefficient,
\begin{equation}
    C^{\rm inter}(\omega,\eta_z;\Lambda) = -\frac{g}{12\pi^2}\ln\frac{\Lambda}{\mu} + C^{\rm fin}(\omega,\eta_z),
    \label{eq:inter-uvlog}
\end{equation}
where $\Lambda$ is the ultraviolet scale (in a lattice, the Brillouin-zone boundary $\Lambda_{\rm BZ}\sim\pi/a$). The ultraviolet log grows slowly [$(g/12\pi^2)\ln(\Lambda/\mu)\approx0.017\ln(\Lambda/\mu)$ at $g=2$], so $|C^{\rm inter}|$ stays below $\sim0.14$ for $\Lambda\le10^3\mu$.

\begin{sloppypar}
    The coefficient $4/3$ is the angular integral $\int_{-\pi/2}^{\pi/2}\cos^3\theta\,d\theta$ of the coherence-factor angular average $\int(1+E\cos\theta)\cos^2\theta\,d\theta=\pi/2+(4/3)E$. \emph{Remark (geometric identification).} One can identify the $4/3$ with the angular integral of the band-geometric tensor $g_{zz}=|\langle u_{-,\mathbf k}|\partial_{k_z}u_{+,\mathbf k}\rangle|^2=u^2/(4E^4)$ over the physical half-plane $k_\perp\ge0$ (Fubini--Study element; see Ref.~\cite{liu2024} for a review): in polar coordinates $u=E\cos\theta$ the metric contributes $\cos^2\theta$ and the far-field phase-space correction contributes $\cos\theta$, giving $\int\cos^3\theta=4/3$. This identification is illustrative only and is not used in any quantitative result: because the PTNR tilt enters through the identity ($\eta_z k_z I$), the band eigenstates and the geometric tensor are $\eta_z$ independent throughout the zone (geometric blindness). The ultraviolet log coefficient $-(g/12\pi^2)$ is therefore strictly $\eta_z$ independent (asymptotic statement: it requires the logarithmic region $E\gtrsim\mu/(1-|\eta_z|)$ to lie inside the cutoff, and it is formulated for a spherical cutoff). At an anisotropic grid a fraction of what a spherical decomposition would book into the log is absorbed into $C^{\rm fin}(\omega,\eta_z;\Lambda_z,\Lambda_\perp)$. 
\end{sloppypar}

The $q^2$ power follows from the coherence factor $F_{+-}\propto q^2$ of Eq.~\eqref{eq:Fpm}. The Fermi-surface-reduced diagnostic $P_2^{\rm FS}$ defined below Eq.~\eqref{eq:P2def} carries the same $q^2$ slope with an analytic coefficient $C^{\rm FS}\equiv P_2^{\rm FS}/q_z^2=-g/16\pi$ at $\eta_z=0$; it is close to---but distinct from---the full occupied-volume coefficient $C^{\rm inter}$ of Eq.~\eqref{eq:P2scaling}: the two are different objects (the Fermi-surface diagnostic is a shell integral that omits the energy denominator of the full bubble, Eq.~\eqref{eq:P2def} versus Eq.~\eqref{eq:P2scaling}; see Sec.~\ref{sec:interband-note}), and $P_2^{\rm FS}$ serves only as a scaling diagnostic---it is not used in the plasmon analysis.

The imaginary part of $C^{\rm inter}$ vanishes below the interband absorption threshold. From the $\delta$-function pole at $\omega=2E_{\mathbf k}$ combined with the Pauli factor, absorption requires a state with $\eta_z k_z+E_{\mathbf k}>\mu$ and $E_{\mathbf k}=\omega/2$; the minimum such frequency is
\begin{equation}
    \omega_c = \frac{2\mu}{1+|\eta_z|},
    \label{eq:thresh}
\end{equation}
so $\operatorname{Im}C^{\rm inter}(\omega)=0$ for $\omega<\omega_c$. The threshold decreases with tilt, reaching $\mu$ as $\eta_z\to1^-$. This $\omega_c$ is the $q\to0$ interband threshold; at finite momentum transfer the interband edge $\omega_{\min}^{\rm inter}(q)$ of Sec.~\ref{sec:damping-window} decreases with $q$, and the latter is the object that bounds the finite-$q$ damping window. The subscript convention is stressed: $\omega_p$ always denotes the RPA plasmon frequency [Eq.~\eqref{eq:wpcorr}], while $\omega_c$ is this absorption threshold---the two are compared explicitly in Sec.~\ref{sec:plasmons}. The finite-$q$ wavevector at which the self-consistent mode meets this edge is listed in Table~\ref{tab:qinter} and shown in Fig.~\ref{fig:window} (Sec.~\ref{sec:damping-window}).

\subsection{Relation to earlier wave-vector-scaling results}

The $q^2$ scaling at the untilted point ($\eta_{\rm RK}=\pi/2$) is already present in the Rhim--Kim RPA~\cite{rhim2016}, where the polarizability reduces to the graphene-like form $P_R=gk_0P_G$ with a $q^2$ coefficient. What is new in this work is the extension to the tilted case and the identification of the UV-logarithmic coefficient $C^{\rm inter}$ and its $\eta_z$ independence. The $q^2$ law obtained here is the standard long-wavelength scaling of the intraband and interband contributions, which is also the reference behavior in the recent study of Ref.~\cite{pandey2025} for three-dimensional Dirac nodal-line systems: the new element identified there is a \emph{resonant} interband term with cubic wave-vector dependence, appearing in the long-wavelength limit when the chemical potential approaches the band edge, in contrast to the quadratic dependence of the standard processes. That resonance belongs to a different regime from the doped one considered here ($q_z\ll\mu/(1+|\eta_z|)$, with the Fermi level away from the band edge), so the two limits are not interchangeable: the coefficient $C^{\rm inter}$ derived above is the standard $q^2$ one, evaluated at a frequency below the threshold $\omega_c$ of Eq.~\eqref{eq:thresh}, where it is weakly frequency dependent (Sec.~\ref{sec:interband}).

\subsection{Correction to the plasmon frequency}

\begin{figure}[t]
    \centering
    \includegraphics[width=0.8\columnwidth]{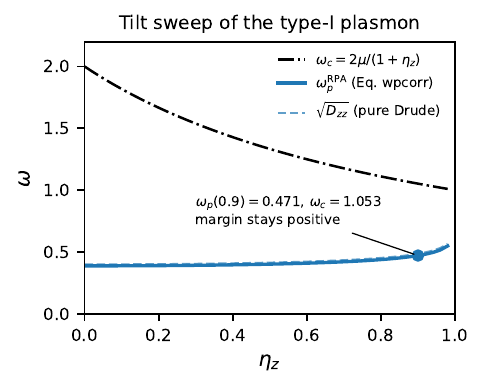}
    \caption{Tilt sweep of the type-I collective mode. $D_{zz}(\eta_z)$ is evaluated from the exact
    reduced integral (Eq.~\eqref{eq:Dzz_reduced}; anchors $g/(4\pi)=0.1592$ at $\eta_z=0$, $0.1708$ at
    $0.5$, $0.2351$ at $0.9$),
    $\omega_p^{\rm RPA}=\sqrt{D_{zz}/(1-C^{\rm inter})}$ at $g_c=1$ with the $E_{\max}=10$ value $C^{\rm inter}=-0.0584$ (the weak $\eta_z$ dependence, below $0.5\%$ over the range shown, is carried by $C^{\rm fin}$; Sec.~\ref{sec:interband}), and the interband threshold $\omega_c=2\mu/(1+\eta_z)$ falls with tilt. At $q\to0$ the margin
    $\omega_c-\omega_p$ remains positive across the type-I range scanned ($\eta_z\le0.98$), consistent
    with the kinematic analysis of Sec.~\ref{sec:damping-window}; at finite $q$ the mode disperses upward and the damping boundary is instead set by the interband edge (Sec.~\ref{sec:damping-window}).}
    \label{fig:tiltsweep}
\end{figure}

Within the random-phase approximation, in the same dielectric-function convention used for nodal-line semimetals in Refs.~\cite{rhim2016,yan2016} and, for Dirac matter generally, in the random-phase-approximation screening and plasmon literature for the massless and massive Dirac plasma~\cite{dsssarma2009,hwang2007,sachdeva2015,thakur2017} and the Dirac--Weyl liquids~\cite{hofmann2015}, the dielectric function is $\varepsilon(q,\omega)=1-V(q)\Pi(q,\omega)$ with $V(q)=4\pi e^2/(\epsilon_b q^2)$ and $\Pi=\Pi_{\rm intra}+\Pi_{\rm inter}$. The long-wavelength intraband (Drude) channel dominates, $\Pi_{\rm intra}\approx D_{zz}q^2/\omega^2$, and the interband channel contributes $\Pi_{\rm inter}\approx C^{\rm inter}(\omega)\,q^2$. The plasmon condition $\varepsilon(q,\omega_p)=0$ then gives the multiplicative correction (with $V=4\pi e^2/\epsilon_b$ in our units $q^2V(q)=V$)
\begin{equation}
    \omega_p^2(\eta_z) = \frac{4\pi e^2}{\epsilon_b}\,
    \frac{D_{zz}(\eta_z)}{1-\frac{4\pi e^2}{\epsilon_b}C^{\rm inter}(\omega_p)},
    \label{eq:wpcorr}
\end{equation}
which differs from the additive form $D_{zz}+C^{\rm FS}q^2$ that would follow from treating the interband channel as an independent additive correction; the multiplicative structure is the RPA consequence of $\Pi_{\rm inter}\approx C^{\rm inter}q^2$ entering the same dielectric function as the Drude channel. It is convenient to introduce the dimensionless coupling $g_c\equiv4\pi e^2/(\epsilon_b\hbar v_0)$ (so that $g_c=4\pi e^2/\epsilon_b$ in the units $\hbar=v_0=1$ of Sec.~\ref{sec:model}), in terms of which
\begin{equation}
    \omega_p^2(\eta_z) = g_c\,\frac{D_{zz}(\eta_z)}{1-g_c\,C^{\rm inter}(\omega_p)}.
    \label{eq:wpcorr_gc}
\end{equation}
In the present units $g_c=1$ (i.e.\ $\alpha_c\equiv g_c/4\pi=1/4\pi\approx0.08$), so $|g_c C^{\rm inter}|\approx0.06\ll1$ and the interband correction to the plasmon frequency is $\lesssim3\%$ at the numerical cutoff, i.e.\ the interband channel is subleading (the slow ultraviolet growth of Eq.~\eqref{eq:inter-uvlog} keeps the correction at the few-percent level up to $\Lambda\sim10^3\mu$). The mode-frequency argument of $C^{\rm inter}$ is immaterial at this level. The dimensionless coupling in real nodal-line materials, however, can be much larger: order-of-magnitude estimates for nodal-line and related semimetal compounds span $\alpha_c\sim0.2$--$5$ ($g_c\sim2.5$--$63$)~\cite{xie2015,topp2017}, while representative nodal-line materials sit at the low-$\alpha_c$ end, so the RPA correction can range from a few percent to order unity and the perturbative regime $|g_c C^{\rm inter}|\ll1$ is not guaranteed. The long-wavelength plasmon is free of interband damping only while $\omega_p(g_c)<\omega_c(\eta_z)$; restoring the coupling, this condition reads $g_c D_{zz}/(1-g_c C^{\rm inter})<[2\mu/(1+|\eta_z|)]^2$ and fails above a tilt-dependent threshold $g_c^{(c)}(\eta_z)$. At $\eta_z=0.9$ the threshold is $g_c^{(c)}\approx6.5$ ($\alpha_c\approx0.5$), so the undamped-window statement of Sec.~\ref{sec:plasmons} holds in the present units ($g_c=1$) but would be reversed for materials with $g_c\gtrsim g_c^{(c)}$. Figure~\ref{fig:tiltsweep} sweeps the type-I tilt: $\omega_p^{\rm RPA}(\eta_z)=\sqrt{D_{zz}(\eta_z)/(1-C^{\rm inter})}$ rises with $\eta_z$ while the threshold $\omega_c=2\mu/(1+\eta_z)$ falls, the margin $\omega_c-\omega_p$ staying positive across the accessible type-I range.

All quantitative plasmon statements in this paper are therefore made at the reference coupling $g_c=1$ and should be rescaled via Eq.~\eqref{eq:wpcorr_gc} before comparison with a specific material.

\subsection{Intrinsic limit and validity of the $q^2$ expansion}

In the intrinsic limit $\mu\to0$, the small-$q$ expansion window closes: the coherence-factor expansion, Eq.~\eqref{eq:Fpm}, requires $q_z\ll E_{\mathbf k}$ over the contributing region, which is controlled by $\mu$ (the states with $E_{\mathbf k}\gtrsim\mu/(1+|\eta_z|)$ dominate), so the $q^2$ coefficient remains ultraviolet cutoff dependent as in Eq.~\eqref{eq:inter-uvlog}, and the interband absorption threshold $\omega_c=2\mu/(1+|\eta_z|)$ vanishes, opening interband damping at all $\omega>0$. This regime is outside the present doped analysis. For the doped case, the small-$q$ expansion of the coherence factor requires $q_z\ll\mu/(1+|\eta_z|)$.

\subsection{Relation to the Fermi-surface-reduced estimate}
\label{sec:interband-note}

For completeness we clarify the relation between Eq.~\eqref{eq:P2def} and the Fermi-surface-reduced expression $P_2^{\rm FS}=-g\int d^3k\,F_{+-}\,\delta(\mu-\varepsilon_{\mathbf k,+})$. The latter selects the occupied conduction-band boundary and omits the energy denominator of Eq.~\eqref{eq:P2def}; it is a boundary contribution to the small-$q$ expansion, not the full interband coefficient. In this paper the interband correction to the plasmon uses the full $C^{\rm inter}$ of Eq.~\eqref{eq:P2scaling}; the Fermi-surface-reduced $P_2^{\rm FS}$ appears only as a scaling diagnostic (its $q^2$ slope is the same).

\section{Plasmon dispersion and collective response}
\label{sec:plasmons}

Given the long-wavelength plasmon frequency of Sec.~\ref{sec:interband} [Eq.~\eqref{eq:wpcorr}], the collective response of the tilted NLSM has two further aspects: the tilt-driven anisotropy of the plasmon frequency and the finite-wavevector dispersion. In the untilted three-dimensional nodal-line model the long-wavelength plasmon frequency follows the $\omega_p\propto n^{1/4}$ density law~\cite{yan2016}; the dispersive~\cite{losic2022}, imbalanced~\cite{islam2021} and nonsymmorphic~\cite{cao2023} variants show how interband physics and lattice dispersion modify that simple law, and the tilt analyzed in this paper adds two inequivalent Drude channels on top of it. The anisotropy follows directly from the two Drude laws of Secs.~\ref{sec:logenh}--\ref{sec:transverse}. The longitudinal and in-plane plasmon frequencies share the multiplicative interband structure of Eq.~\eqref{eq:wpcorr} but inherit different Drude weights; the interband coefficients enter each channel as $O(|VC^{\rm inter}|)\lesssim6\%$ corrections. The transverse interband coefficient $C^{xx}$ (momentum transfer in the plane) is smaller than the longitudinal one, $C^{xx}\approx-0.023$ versus $C^{\rm inter}\approx-0.0584$ at the numerical cutoff (Appendix~\ref{app:interband}), so in the ratio the interband factor is $(1-VC^{xx})/(1-VC^{\rm inter})\xrightarrow{V=1}0.966$---a few-percent ($\sim3.4\%$) suppression, not a cancellation---and the leading-order anisotropy is set by the Drude weights,
\begin{equation}
    \frac{\omega_{p,\perp}^2}{\omega_{p,\parallel}^2}
    \sim \frac{D_\perp}{D_{zz}} \sim \frac{1}{\delta\ln(1/\delta)}
    \label{eq:aniso_ratio}
\end{equation}
which diverges as the type-I/II transition~\cite{soluyanov2015,tan2022} is approached from the type-I side ($\delta\to0^+$) within the tail-inside-cutoff window of Sec.~\ref{sec:transverse} (i.e.\ $\delta\gg1/\Lambda$ with $\Lambda$ the physical cutoff); at fixed physical cutoff the ratio saturates at the cutoff-bound plateau of Sec.~\ref{sec:transverse}, so the divergence is a window statement, not an unconditional one.
Equation~\eqref{eq:aniso_ratio} is the leading asymptotic form as $\delta\to0$ and is not accurate as a numerical prediction over the accessible tilt range: evaluating the exact reduced
integrals of Secs.~\ref{sec:logenh}--\ref{sec:transverse} at $\mu=k_0$ gives $D_\perp/D_{zz}=1.05$, $1.19$, $1.53$ and $3.04$ at $\eta_z=0.3$, $0.5$, $0.7$ and $0.9$, whereas the
asymptotic form would give $4.0$, $2.9$, $2.8$ and $4.3$; the anisotropy therefore grows monotonically with tilt, but its quantitative value must be taken from the exact integrals
rather than from Eq.~\eqref{eq:aniso_ratio}. As discussed in Sec.~\ref{sec:interband}, in the doped type-I regime the long-wavelength plasmon is not damped by interband processes; the tilt lowers the absorption threshold $\omega_c=2\mu/(1+|\eta_z|)$ toward $\mu$ as $\eta_z\to1^-$, bringing the $q\to0$ mode and the interband continuum closer (the $q\to0$ margin stays positive for all accessible type-I tilts). At finite $q$, however, the mode disperses upward, and Sec.~
\ref{sec:damping-window} shows that it crosses the interband continuum edge at the self-consistent entry
$q_{\rm inter,self}(\eta_z)$ (Fig.~\ref{fig:window}): the channel that opens with increasing $q$ is the interband
one, while the intraband channel stays closed for $q\lesssim0.8$. A quantitative damping rate across the type-I/II transition, which requires the type-II side outside the present weak-tilt expansion, is left for future work. The quantitative type-II plasmon dispersion is cutoff dependent through $D_{zz}$; we report type-II values as representative at fixed cutoff, and do not generalize them into universal type-II physics. For candidate nodal-line materials (e.g.\ the Ca$_3$P$_2$ family of Ref.~\cite{xie2015} or transition-metal nodal-line compounds), the tilt $\eta_z$ is not a material constant but a band-structure asymmetry parameter to be estimated from first principles; the present analysis treats it as a free parameter and predicts the anisotropy and threshold trends as functions of it.

\begin{figure}[t]
    \centering
    \includegraphics[width=0.85\columnwidth]{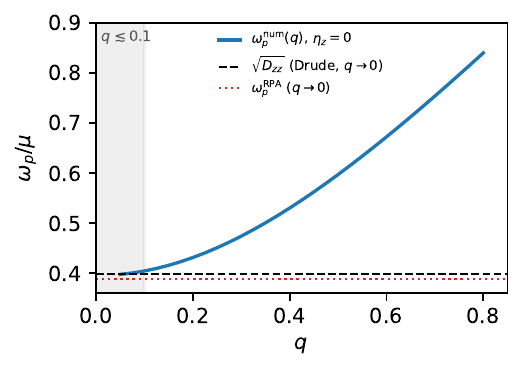}
    \caption{Long-wavelength plasmon frequency at $\eta_z=0$: the self-consistent mode $\omega_p^{\rm num}(q)$ (solid), the pure-Drude value $\omega_p=\sqrt{D_{zz}}=0.399$ (dashed, $q\to0$), and the RPA value $\omega_p=\sqrt{D_{zz}/(1-C^{\rm inter})}=0.388$ (dotted), which includes the interband correction (a $2.8\%$ effect). Within the small-$q$ window $q\lesssim0.1$ (shaded) the multiplicative structure keeps $\omega_p$ within a few percent of its $q\to0$ value; at this tilt the interband threshold $\omega_c=2\mu$ lies well above the mode, so the long-wavelength plasmon is undamped.}
    \label{fig:plasmon}
\end{figure}

The crossover wavevector $q_*$ at which the interband correction to the $q$-dependent part of the gap would equal the Drude contribution is not a meaningful scale in this system: because the interband coefficient $C^{\rm inter}$ is independent of $q$ to leading order [Eq.~\eqref{eq:P2scaling}], the multiplicative correction of Eq.~\eqref{eq:wpcorr} does not introduce a $q$-dependent competition with $D_{zz}$. The interband channel therefore does not dominate at any wavevector within the validity range $q_z\lesssim0.1$ of the small-$q$ expansion, contrary to the picture in which the two channels enter additively at a scale $q_*$.

\subsection{Landau damping and the self-consistent plasmon window}
\label{sec:damping-window}

The long-wavelength plasmon is free of Landau damping only while its frequency lies outside the particle--hole
continuum. The relevant object is the \emph{support} of the continuum with its spectral weight,
\begin{equation}
    \operatorname{supp}_{ss'}(q)=\overline{\Bigl\{\Delta\varepsilon_{ss'}(\mathbf k,\mathbf q):\
    \bigl[f_s(\mathbf k)-f_{s'}(\mathbf k+\mathbf q)\bigr]\bigl|F_{ss'}\bigr|^2\ne0\Bigr\}},\qquad
    \operatorname{supp}=\bigcup_{ss'}\operatorname{supp}_{ss'},
    \label{eq:supp}
\end{equation}
where $\Delta\varepsilon_{ss'}=\eta_z q_z+s'E_{\mathbf k+\mathbf q}-sE_{\mathbf k}$ and the weight includes the
coherence factor, whose value at the intraband edge is unity. Defining the support by
the weighted set of Eq.~\eqref{eq:supp} is essential: extremizing $\Delta\varepsilon_{-+}$ over all $\mathbf k$
would give a vanishing threshold at $q\to0$, because the nodal ring has $E_{\mathbf k}\to0$---but those states lie
\emph{inside} the Fermi surface, where the Pauli factor vanishes; with the weight included the interband edge
tends to $\omega_c$ of Sec.~\ref{sec:interband}.

On the Fermi surface ($E_{\mathbf k}=\mu-\eta_z k_z$), the longitudinal group velocity
$v^{(+)}_z=\eta_z+k_z/E_{\mathbf k}$ increases strictly with $k_z$ [$dv_z/dk_z=\mu/E^2>0$], so its extrema sit at
the endpoints of the allowed interval $k_z=\pm\mu/(1\pm\eta_z)$:
\begin{equation}
    v_z^{\max}=1+\eta_z,\qquad v_z^{\min}=\eta_z-1,\qquad
    |v_x|^{\max}=1 \quad(\text{for }\mathbf q\perp\hat z),
    \label{eq:vmax}
\end{equation}
with the transverse maximum attained at $k_z=0$ and strictly tilt independent. Equation~\eqref{eq:vmax} turns the
type-I/II criterion into a kinematic statement: the longitudinal window closes, $v_z^{\min}=0$, precisely at
$\eta_z=1$. For $\mathbf q\parallel\hat z$ the positive-frequency intraband continuum therefore occupies
$[0,(1+\eta_z)q]$ in the type-I regime. If the mode frequency were independent of $q$, the mode would remain
above this continuum only for
\begin{equation}
    q<q^*_{\parallel}\simeq\frac{\omega^{(0)}_{p,z}}{1+\eta_z}\quad(\mathbf q\parallel\hat z),
    \label{eq:qstar}
\end{equation}
where $\omega^{(0)}_{p,z}$ is the $q\to0$ mode frequency of Sec.~\ref{sec:plasmons}. Equation~\eqref{eq:qstar} is
a \emph{long-wavelength estimate}: it uses the $q\to0$ mode frequency, whereas the mode disperses upward with $q$.
It is therefore reliable only inside the small-$q$ window $q\lesssim0.1$ in which the multiplicative RPA structure
of Eq.~\eqref{eq:wpcorr} keeps $\omega_p$ $q$-independent to within a few percent (Fig.~\ref{fig:plasmon}). Being built from the
$q\to0$ frequency alone, however, the estimate cannot decide whether the intraband channel actually opens at
finite $q$.

To answer that question we evaluate the mode position self-consistently at finite $q$. The mode frequency
$\omega_p^{\rm num}(q)$ is the zero of $\mathrm{Re}\,\varepsilon(q,\omega)$ obtained by a global frequency sweep
followed by root refinement, taking the zero closest to the maximum of the loss function
$-\mathrm{Im}\,\varepsilon^{-1}(q,\omega)$; the continuum edges are computed with the full spectral weight of
Eq.~\eqref{eq:supp} (Appendix~\ref{app:window}). The calculation covers $q\in[0.05,0.8]$ at
$\eta_z=0,\,0.3,\,0.5,\,0.7,\,0.9$. The result contradicts the estimate-based picture: the mode \emph{rises} with
$q$ and remains \emph{above} the intraband upper edge $(1+\eta_z)q$ over the entire studied range---e.g.\
$\omega_p^{\rm num}(q=0.5)=1.001$ versus $(1+\eta_z)q=0.95$ and $\omega_p^{\rm num}(q=0.7)=1.344$ versus $1.33$
at $\eta_z=0.9$. The margin $\omega_p^{\rm num}(q)-(1+\eta_z)q$ decreases monotonically over the scan, from $+0.372$
at $q=0.05$ to $+0.0069$ ($0.5\%$ of the edge value) at $q=0.775$, and remains positive at every $q$ at which
the mode is resolved (no zero of $\mathrm{Re}\,\varepsilon$ is found at $q=0.8$, where the mode has merged with
the interband continuum); the intraband
channel is therefore not entered at any resolved grid point of the scan, and the conclusion is correspondingly sensitive
to the grid and cutoff at the level of this margin. The intraband channel therefore \emph{does not open} in the resolved scan ($q\lesssim0.8$): the intraband
boundary predicted by Eq.~\eqref{eq:qstar} is not realized as an actual boundary there, because the estimate
extrapolates the $q\to0$ frequency outside the small-$q$ window in which it is valid.

\begin{figure}[t]
    \centering
    \includegraphics[width=\columnwidth]{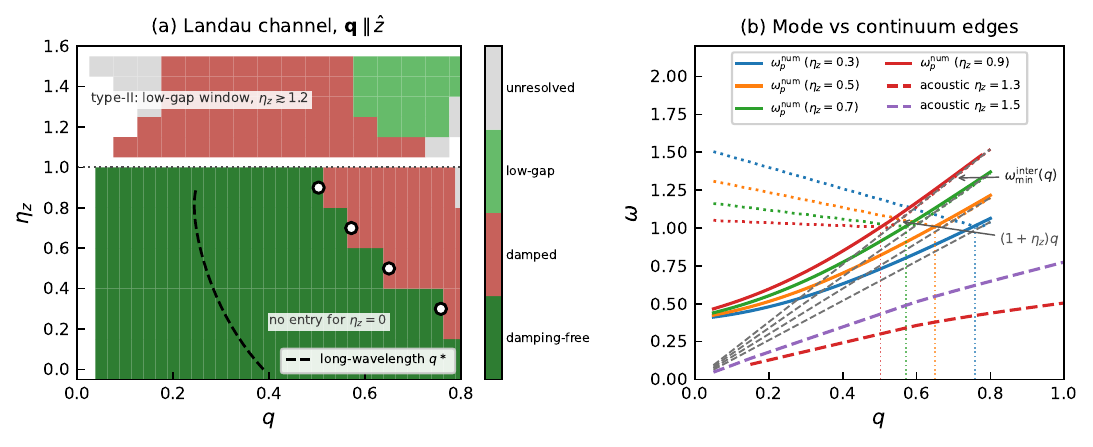}
    \caption{(a) Landau channel in the $(q,\eta_z)$ plane for $\mathbf q\parallel\hat z$, classified with the
    self-consistent mode position $\omega_p^{\rm num}(q)$: free of Landau damping (the mode lies between the
    intraband upper edge and the interband edge) or above the interband edge (damped). The intraband channel does not open for $q\lesssim0.8$; the damping-free window is bounded by the interband
    entry. The mode remains above the intraband continuum $[0,(1+\eta_z)q]$ at every resolved grid point, so no
    intraband-damped region is identified in the scan. Filled dots mark the self-consistent interband entry $q_{\rm inter,self}=0.76,\,0.65,\,0.57,\,0.50$ at
    $\eta_z=0.3,\,0.5,\,0.7,\,0.9$. The dashed line is the long-wavelength estimate of Eq.~\eqref{eq:qstar}, which is not
    realized as a boundary beyond the small-$q$ window in which it is valid. In the type-II regime ($\eta_z>1$) the low-gap window (green filled region) opens for
    $\eta_z\gtrsim1.2$ at $q\gtrsim q_{\rm low}(\eta_z)\approx0.6$ (Table~\ref{tab:qinter-typeII}; the
    boundary position is grid-limited below $\eta_z\approx1.2$); at
    $\eta_z=1.1$ the acoustic branch is not resolved. (b) Self-consistent
    mode dispersion $\omega_p^{\rm num}(q)$ (solid; the finite-$q$ zero of $\mathrm{Re}\,\varepsilon(q,\omega)$, to be distinguished from the $q\to0$ analytic frequency $\omega_p^{\rm RPA}$) against the intraband upper edge $(1+\eta_z)q$ (dashed gray) and
    the interband edge $\omega_{\min}^{\rm inter}(q)$ (dotted); the mode rises with $q$ above the intraband edge
    and crosses the interband edge at $q_{\rm inter,self}$ (dotted vertical lines), above which it becomes subject
    to interband Landau damping. In the type-II regime the low-frequency acoustic branch (green) enters the low-gap window at
    $q\gtrsim q_{\rm low}$.}
    \label{fig:window}
\end{figure}

The channel that actually bounds the window is therefore the interband one. Its edge $\omega_{\min}^{\rm inter}(q)$
decreases with $q$ (the finite momentum transfer opens additional phase space), and the self-consistent mode
crosses it at the entries of Table~\ref{tab:qinter}; the intersection
$\omega_p^{\rm num}(q)=\omega_{\min}^{\rm inter}(q)$ is shown in Fig.~\ref{fig:window}. For $\eta_z=0$ the mode
stays below the interband edge over the whole range $q\le0.8$, so no interband entry exists there. The
At $\eta_z=0.9$ the tail end $-\mu/\delta=-10$ approaches the cutoff, so the entry is evaluated with the bubble
inside the hard cutoff $\Lambda_z=12$ of Sec.~\ref{sec:ward}, i.e.\ inside the same tail-inside-cutoff condition
that underlies the logarithmic window of Sec.~\ref{sec:logenh}: with a box of half-width $6$ the numerical mode
at this tilt fails to reproduce the analytic $q\to0$ frequency $\omega_p^{\rm RPA}=0.471$, whereas $\Lambda_z=12$
gives $\omega_p^{\rm num}(q=0.05)=0.467$. The low-$q$ value is stable under refinement: raising $\Lambda_z$ to
$16$ and $20$ at fixed $k_z$ spacing changes it by $7\times10^{-4}$ and then $1.5\times10^{-5}$, refining the $k$
grid from $1000\times2000$ to $2000\times4000$ changes it by $4\times10^{-4}$, and reducing the broadening from
$5\times10^{-3}$ to $1.25\times10^{-3}$ changes it by $8\times10^{-5}$. The interband channel is
therefore closed only for $q<q_{\rm inter,self}(\eta_z)$, with $q_{\rm inter,self}$ decreasing as the tilt
approaches unity.
\begin{table}[t]
    \centering
    \caption{Self-consistent interband entry $q_{\rm inter,self}$ at which the numerical mode
    $\omega_p^{\rm num}(q)$ crosses the interband edge $\omega_{\min}^{\rm inter}(q)$ (the scan of
    Sec.~\ref{sec:damping-window} uses a uniform grid of spacing $\Delta q=0.025$ over $q\in[0.05,0.8]$, and
    each entry is localized within a bracketing interval of width $\Delta q$, the quoted value being obtained
    by linear interpolation of the two edge curves over that interval; Appendix~\ref{app:window}). For the four tabulated nonzero tilts the
crossings occur at $q_{\rm inter,self}\simeq0.95$--$0.99\,\mu/(1+|\eta_z|)$, i.e.\ near the nominal breakdown scale of
the small-$q$ expansion of Sec.~\ref{sec:interband}; they are obtained from the full bubble of
Appendix~\ref{app:window}, not from the expansion, evaluated inside the hard cutoff $\Lambda_z=12$
that contains the Fermi-surface tail (Sec.~\ref{sec:damping-window}). Spot checks with $\Delta q$ reduced to
$0.0025$ and with the cutoff, the $k$ grid and the broadening varied leave the entries of the table unchanged to
within $0.002$. For $\eta_z=0$ the mode remains below the interband edge for all $q\le0.8$ and no entry exists.}
    \label{tab:qinter}
    \begin{ruledtabular}
        \begin{tabular}{lccccc}
            $\eta_z$ & 0 & 0.3 & 0.5 & 0.7 & 0.9 \\
            \hline
            $q_{\rm inter,self}$ & --- & $0.76$ & $0.65$ & $0.57$ & $0.50$ \\
        \end{tabular}
    \end{ruledtabular}
\end{table}

\begin{figure}[t]
    \centering
    \includegraphics[width=0.62\columnwidth]{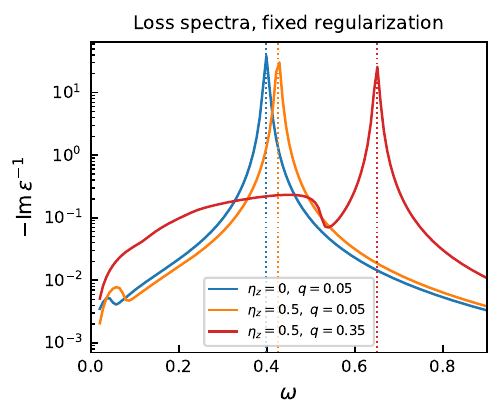}
    \caption{Loss function $-\mathrm{Im}\,\varepsilon^{-1}(q,\omega)$ at three representative points
    $(q,\eta_z)=(0.05,0)$, $(0.05,0.5)$, and $(0.35,0.5)$, all in the damping-free region under the
    classification of Fig.~\ref{fig:window}, at fixed regularization (hard-cutoff box $\Lambda_z=4$, $\Lambda_\perp=3$, thermal smearing $T=5\times10^{-3}$); dotted vertical lines mark the zeros of $\mathrm{Re}\,\varepsilon(q,\omega)$. There the peak coincides with the mode zero in the weak-damping regime (the operational criterion $\Gamma_{\rm eff}/\omega_{\rm peak}\lesssim0.3$ of Appendix~\ref{app:window}); at $q=0.35$ the mode remains a well-defined resonance, its upward shift relative to $q\to0$ following the finite-$q$ mode dispersion of Fig.~\ref{fig:window}. The low-level tail inside the window is a kernel-broadening artifact; the strict statement $\Gamma\equiv0$ follows from the un-broadened support criterion. These spectra are illustrative and no universal damping rate is extracted
    from them.}
    \label{fig:lossspectra}
\end{figure}

The corresponding spectra are shown in Fig.~\ref{fig:lossspectra}. In the type-II regime the kinematics change qualitatively: $v_z^{\min}=\eta_z-1>0$ opens a low-frequency gap in
the intraband continuum, $[(\eta_z-1)q,(1+\eta_z)q]$ (with the weighted support edge of Eq.~\eqref{eq:supp} larger than the bare kinematic value; see below). Self-consistent evaluation at finite $q$ (global
frequency sweep with the weighted support edges, $q\in[0.05,1.00]$ at $\eta_z=1.1$--$1.5$) reveals two
collective branches: a high-frequency optical branch ($\omega\sim1.4$--$2.1$) that lies above the interband
edge and is inter-band damped at all $q$, and a low-frequency acoustic-like branch ($\omega\sim0.5q$) that
lies below both the intraband lower edge and the interband edge for $q\gtrsim q_{\rm low}(\eta_z)$ and
$\eta_z\gtrsim1.2$ (Table~\ref{tab:qinter-typeII}). The branch is reported as a \emph{candidate}: its identity rests on the loss-peak-closest criterion alone, its slope stays close to $0.5q$ while the tilt changes $v_z^{\min}$ by a factor of about $2.5$, and no independent cross-check (pole tracking under a change of regulator, or a residue analysis) has been performed, so we do not claim it as an established collective mode.

The branch gives rise to a \emph{low-gap damping-free window} that opens as the tilt increases beyond unity. The window is the type-II analogue of the gapless undamped
plasmon in tilted Dirac semimetals~\cite{sadhukhan2020}, akin to the infrared plasmons that propagate through a hyperbolic nodal metal~\cite{shao2022}, here realized on the open Fermi-surface pockets of
the tilted nodal ring; the correspondence is quantitative in kind but hardened in threshold---in the tilted
Weyl case the undamped mode appears throughout the type-II regime, whereas the nodal-ring geometry (two radial
roots and the weighted support edges) requires $\eta_z\gtrsim1.2$ before the acoustic branch clears both edges. At $\eta_z=1.1$ the acoustic branch is not resolved in the present $q$ grid (the loss
spectrum is dominated by the inter-band damped optical branch at all $q$), so whether a window opens at the
smallest type-II tilts remains an open question; no known nodal-line compound is established to reach type-II tilts, so the
low-gap window is currently a theory-side prediction with the tilt to be surveyed compound-by-compound from
first principles. All type-II mode frequencies are reported at fixed regularization and are not claimed as universal type-II physics. The window itself is not an artifact of the reference cutoff: recomputing the in-window zeros of $\mathrm{Re}\,\varepsilon$ with the longitudinal cutoff varied over $\Lambda_z=4$--$16$ at fixed $\Lambda_\perp=4$ leaves them present at every cutoff for $\eta_z\ge1.3$ (and at $\Lambda_z\ge6$ for $\eta_z=1.2$), with the mode frequency maintaining its $q$ linearity.

Two caveats on the type-II domain are stated for completeness: (i) the numerical type-II integrals retain only the $s=+$ open-pocket sheet, whereas at $\eta_z>1$ the $s=-$ band acquires a second Fermi sheet inside the cutoff for $k_z\ge\mu/(\eta_z-1)$ (e.g.\ $k_z\gtrsim2.0$ at $\eta_z=1.5$; the shell condition $E=\eta_z k_z-\mu\ge|k_z|$ with $k_z>0$ gives this threshold). A dedicated evaluation of the omitted sheet on the $k$-space grid gives its share of the platform weight as $0.03\%$, $0.03\%$, $0.06\%$, $10\%$ and $19\%$ at $\eta_z=1.1$, $1.2$, $1.3$, $1.4$ and $1.5$ respectively: for $\eta_z\le1.3$ the second sheet lies beyond the cutoff and the omission is immaterial, while at $\eta_z=1.5$ the quoted type-II weights are lower bounds by about $19\%$; (ii) the type-II Fermi surface is open in $k_\perp$ as well, so the hard-cutoff box does not fully contain it and the boundary-term argument of Sec.~\ref{sec:ward} is not asserted for the type-II platform.

In the type-II regime the Fermi surface is open and $k_z$ extends to $-\infty$ in the thermodynamic limit; in a finite system the Brillouin-zone cutoff $|k_z|\le\pi$ regularizes the divergence, and all type-II results are quoted at this BZ-regularization platform, a scheme-dependent choice: no universal (scheme-independent) type-II statement is claimed for the quantitative Drude weight or the collective response built on it.
\begin{table}[t]
    \centering
    \caption{Low-gap damping-free window threshold $q_{\rm low}(\eta_z)$ for type-II tilt, below which the
    low-frequency acoustic branch is not resolved (the loss spectrum is dominated by the inter-band
    damped optical branch) and above which it lies in the low-gap window (below both the intraband lower
    edge and the interband edge). For $\eta_z=1.1$ no gap-window is resolved in the present scan, and the $\eta_z\approx1.2$ boundary position is grid-limited (see Appendix~\ref{app:window}). The weighted intraband lower edge (Eq.~\eqref{eq:supp}) is a factor $\sim$2--3
    larger than the bare kinematic value $(\eta_z-1)q$. }
    \label{tab:qinter-typeII}
    \begin{ruledtabular}
        \begin{tabular}{lccccc}
            $\eta_z$ & 1.1 & 1.2 & 1.3 & 1.4 & 1.5 \\
            \hline
            $q_{\rm low}$ & --- & $0.65$ & $0.60$ & $0.60$ & $0.60$ \\
        \end{tabular}
    \end{ruledtabular}
\end{table}

\section{Ward identity and gauge consistency of the response tensor}
\label{sec:ward}

The response tensor is evaluated in two regularization schemes: scheme A (hard-cutoff cylinder $|k_z|\le\Lambda_z$, $k_\perp\le\Lambda_\perp$, $\Lambda_z=12$, $\Lambda_\perp=4$) and scheme C (reference-subtraction, $\eta_z=0$ at identical cutoffs). The full derivation (five-term decomposition, telescoping per-$k$ identity, and the boundary-term theorem) is given in Appendix~\ref{sec:ward-appendix}; we state the results here. The density-longitudinal channel ($q_\mu\Pi^{\mu 0}$) satisfies the Ward identity $i\Omega_m\Pi^{00}(q,i\Omega_m)=\lambda\Pi^{x0}(q,i\Omega_m)$ at momentum transfer below the boundary-crescent validity bound and finite Matsubara frequency within the hard-cutoff domain containing the Fermi surface, because the per-$k$ identity integrates to zero (the boundary term vanishes since the occupation number $N(k)=1$ throughout the crescent edge region). The identity holds in the static and dynamic limits alike, the same boundary cancellation applying below the boundary-crescent validity bound and at finite frequency. The density channel whose consistency is established here is the channel that enters the finite-$q$ dielectric function of the damping-window analysis (Appendix~\ref{app:window}), whose numerical outcome is summarized in Fig.~\ref{fig:window} and Table~\ref{tab:qinter}.

In the transverse channel ($\nu=x$) the contact term $\Gamma^{xx}\neq0$ leaves a frequency-independent residual $R^{xx}(\Lambda_z)$ that grows with the hard cutoff but not linearly---over $\Lambda_z=4$--$40$ it follows $R^{xx}\simeq(7.6\ln\Lambda_z-2.5)\times10^{-3}$, the increment per unit cutoff falling by a factor of seven across that range---and it is linear in the momentum transfer, $R^{xx}/\lambda\simeq0.33$ at $\Lambda_z=12$. The residual is a hard-cutoff artifact that scheme-C subtraction reduces but does not eliminate; it is traced to the contact term and is scheme dependent (Appendix~\ref{sec:ward-appendix}). Full tensor gauge consistency is therefore claimed only for the density-longitudinal channel (Appendix~\ref{sec:ward-appendix}); the boundary cancellation itself is direction-independent (Appendix~\ref{sec:ward-appendix}, \emph{General transfer direction}), the numerical evaluation is carried out at the reference direction $\mathbf q=\lambda\hat x$.

\section{Conclusion}
\label{sec:conclusion}

This paper analyzes the anisotropic response of three-dimensional tilted nodal-line semimetals, showing that the tilt $\eta_z$ controls the Drude weights through distinct tail-region mechanisms and that the finite-wavevector collective response combines an interband correction---IR-integrable by Pauli blocking at finite $\mu$ but carrying a slow ultraviolet logarithm---with gauge-consistent vertices. Within the window where the Fermi-surface tail lies inside the ultraviolet cutoff, the longitudinal Drude weight grows logarithmically ($D_{zz}\sim B\ln(1/\delta)$) because the tail integrand of Eq.~\eqref{eq:eqn11} approaches $(1-2y)^2/y$, i.e.\ the longitudinal weight per unit $k_z$ scales as $1/|k_z|$ in the matching tail, so that the extensive tail measure is integrated against a $1/y$ singularity; the transverse weight follows a different $\delta^{-1}$ power law because its tail integrand $\mathcal W_x$ tends instead to a constant; at fixed physical cutoffs both saturate at the plateaus of Sec.~\ref{sec:transverse}. These are distinct mechanisms, not different facets of a single ``master parameter.''

Our main results are: (i) $B=g/(2\pi)^2$ (at $\mu=k_0=1$ in the units of Sec.~\ref{sec:model}), determined within
the tail-inside-cutoff window (Sec.~\ref{sec:logenh}); (ii) analytic $\sqrt{\delta}$ correction to both Drude weights; (iii) $P_2\propto q^2$ scaling law (as at the untilted point~\cite{rhim2016}), with a UV-logarithmic coefficient $C^{\rm inter}\approx-0.06$ whose $\eta_z$ independence (a blindness of the geometric weight) and multiplicative RPA role are established here, giving a multiplicative subleading correction $\omega_p^2=D_{zz}/(1-C^{\rm inter})$ and an interband absorption threshold $\omega_c=2\mu/(1+|\eta_z|)$ below which the long-wavelength plasmon is not interband-damped; (iv) density-channel Ward identity exact below the boundary-crescent validity bound and at finite frequency within the hard-cutoff domain that contains the Fermi surface; (v) the Landau-damping window of the long-wavelength plasmon: the long-wavelength estimate $q^*=\omega_p^{(0)}/(1+\eta_z)$ for $\mathbf q\parallel\hat z$ decreases with tilt, but a self-consistent finite-$q$ evaluation shows that the mode stays above the intraband continuum, so the intraband channel does not open for $q\lesssim0.8$ and the window is bounded instead by the finite-$q$ interband entry $q_{\rm inter,self}(\eta_z)$ (from $\simeq0.76$ at $\eta_z=0.3$ to $\simeq0.50$ at $\eta_z=0.9$); in the type-II regime ($\eta_z>1$) a low-gap damping-free window opens for $\eta_z\gtrsim1.2$ at $q\gtrsim q_{\rm low}(\eta_z)\approx0.6$ (Table~\ref{tab:qinter-typeII}), the branch responsible being reported as a candidate (Sec.~\ref{sec:damping-window}).

Two practical implications follow. Response calculations in tilted nodal-line semimetals should use transverse line-integral vertices. The interband polarization should be computed from the full Lindhard form with the Pauli factor (Sec.~\ref{sec:interband}); Fermi-surface-reduced estimates capture the scaling but not the full coefficient.

\textbf{Limitations and open questions.} The type-II quantitative results depend on the UV/BZ cutoff choice and are reported at fixed cutoff, not as universal type-II physics. Gauge consistency is established for the density-longitudinal channel only: the transverse channel retains a scheme-dependent hard-cutoff residual that we do not further decompose, and full tensor Ward consistency is not claimed. The interband coefficient is ultraviolet cutoff dependent [Eq.~\eqref{eq:inter-uvlog}] and is quoted at the numerical cutoff of Appendix~\ref{app:interband}; in a lattice it is cut off by the Brillouin zone, and the slow logarithmic growth keeps the correction subleading at the scales studied; exchange--correlation effects beyond the random-phase approximation~\cite{shao2020corr} are not included. In the intrinsic limit $\mu\to0$ the absorption threshold vanishes and the $q^2$ expansion window closes, which is outside the present doped scope.

The logarithmic enhancement $D_{zz}\sim B\ln(1/\delta)$ is likewise a window statement: the logarithmic region must lie inside the UV cutoff ($\delta\gg1/\Lambda$) and the tail must lie inside the Brillouin zone ($\delta\gtrsim k_0/\Lambda_{\rm BZ}$). With a single physical cutoff $\Lambda=\Lambda_{\rm BZ}$ these are the same condition, $\delta\gtrsim k_0/\Lambda$; at $\Lambda\sim\pi k_0$ it confines the analysis to $\delta\gtrsim k_0/\pi\simeq0.32$, where $\ln(1/\delta)$ is at most of order unity. The enhancement of the weight at a fixed lattice cutoff is therefore modest, and the coefficient $B$ is extracted from the continuous-model quadrature (Sec.~\ref{sec:logenh}) rather than from a raw log-slope fit at fixed physical cutoff. A curve-level numerical reproduction of the untilted RPA dispersion of Ref.~\cite{rhim2016} was not attempted. We do not extract a damping rate: within the damping-free region identified in Sec.~\ref{sec:damping-window} (below the interband entry $q_{\rm inter,self}$ in the type-I regime) the Landau channels are kinematically closed, so $\Gamma\equiv0$ at $T=0$, while above the entry the width of the resonance depends on the extraction convention (complex pole versus line shape) and its interband weight inherits the ultraviolet logarithm of $C^{\rm inter}$; we therefore report the window boundary rather than $\Gamma$.

The hard-cutoff regularization is used throughout, with the parameters set by the object at hand: $(\Lambda_z,\Lambda_\perp)=(12,4)$ for the response tensor of Sec.~\ref{sec:ward},
$(\Lambda_z,\Lambda_\perp)=(4,3)$ for the loss spectra and the finite-$q$ mode scan of Sec.~\ref{sec:damping-window}, and the interband cutoff $E_{\max}=10$ for $C^{\rm inter}$
(Appendix~\ref{app:interband}); the continuity structure established in Sec.~\ref{sec:ward} is preserved in the plasmon calculation, and the sensitivity of the mode position to these
choices is at the sub-percent level within the scanned window (Sec.~\ref{sec:damping-window}). Systematic studies of finite-temperature and disordered damping, and of the collective modes across the type-I/II transition, are left for future work, as is a dedicated surface-mode analysis: drumhead-surface plasmons are outside the present bulk treatment, and the bulk window statements apply to the HREELS parameter range where the surface contribution is subdominant. For candidate nodal-line materials (e.g.\ the Ca$_3$P$_2$ family~\cite{xie2015}, the antiperovskite Cu$_3$PdN~\cite{yu2015}, CaAgAs~\cite{wang2017caagw}, and transition-metal nodal-line compounds~\cite{weng2016}), the tilt $\eta_z$ is not a material constant but a band-structure asymmetry parameter to be estimated from first principles; over the tilt range scanned in this work ($\eta_z\le0.9$ in the continuum model; tighter bounds apply if a finite box is imposed, Sec.~\ref{sec:model}), the few-percent plasmon correction and the $D_\perp/D_{zz}$ anisotropy provide target magnitudes (and their tilt trends) for surface-sensitive probes such as HREELS, subject to the material-coupling caveat of Sec.~\ref{sec:plasmons} and to the scheme-dependent transverse-channel residual of Sec.~\ref{sec:ward}, which bounds the accuracy of the $D_\perp/D_{zz}$ anisotropy prediction. A concrete HREELS signature would require resolving the interband entry $q_{\rm inter,self}$ (Table~\ref{tab:qinter}) and the plasmon linewidth within the damping-free region (Sec.~\ref{sec:damping-window}); the required momentum-transfer and energy resolutions are comparable to those of state-of-the-art HREELS~\cite{ekstrom2021,lin2022} or of electron energy-loss spectroscopy of topological semimetals~\cite{song2019,egerton2009} for representative material parameters, and absolute numbers for a specific compound follow once its band velocity and dielectric environment are determined. Complementary probe channels---tilt-resolved Kerr spectroscopy in tilted nodal-loop semimetals~\cite{ekstrom2021} and inelastic-plasmon spectroscopy of nodal-line materials~\cite{lin2022}---would test the anisotropy and threshold trends directly. The division of labour between the two is qualitative but sharp. Kerr spectroscopy operates in the long-wavelength ($q\to0$) limit; the rotation it measures is not magneto-optical---the model is time-reversal invariant and carries no Hall response---but the optical activity of an anisotropic conductor, whose principal axes are rotated with respect to the plane of incidence so that a zero-field Kerr signal tracks the conductivity anisotropy~\cite{ekstrom2021}. In the present geometry (tilt along $z$) the in-plane response remains isotropic, so the anisotropy entering the signal is that between $\sigma_{zz}$ and $\sigma_\perp$, and the quantities tested are the Drude ratio $D_\perp/D_{zz}$ and the interband threshold $\omega_c$---of which $\omega_c$ is the more robust target, because the anisotropy ratio is the quantity bounded by the scheme-dependent transverse residual of Sec.~\ref{sec:ward}. HREELS instead resolves finite momentum transfer and therefore tests the finite-$q$ interband entry $q_{\rm inter,self}$ and the boundary of the damping-free window of the longitudinal (out-of-plane) channel---the in-plane channel instead probes the transverse response, whose finite-$q$ entry is not computed here; because HREELS is surface sensitive, the bulk window statements apply where the bulk loss dominates, and the tilt \emph{trend} of the anisotropy and of the threshold---rather than any single absolute value---is the primary experimental handle, since it requires no material-specific calibration.

\section*{ACKNOWLEDGEMENTS}
The author Wei Li thanks Dr.\ Ipsita Mandal for suggesting this calculation problem and for valuable early-stage correspondence. We are grateful to Prof.\ G.-Z. Liu for the valuable discussions. This work was supported by the Science Research Foundation for High-Level Talents of Anhui University of Science and Technology under Grant YJ20240002 and by the National Natural Science Foundation No. 11304318 and No. 12274414.

\section*{Data Availability Statement}
The data that support the findings of this study and the analysis scripts that generate the figures are available from the corresponding authors upon reasonable request. The analytic results follow from the derivations of the appendices.

\appendix
\section{Reduction of the Kubo formula to one-dimensional Drude integrals}
\label{app:drude}

This appendix records the stepwise reduction of the zero-temperature Kubo Drude formula of Sec.~\ref{sec:drudebench} to the one-dimensional integrals used throughout, and the derivation of the transverse-law crossover function of Sec.~\ref{sec:transverse}. The reduction is exact within the continuum model; no further approximation enters between the Kubo formula and Eq.~\eqref{eq:Dzz_reduced}.

\paragraph{A.1 Band restriction.}
The longitudinal Drude weight is the zero-frequency Kubo weight of the conduction band,
\begin{equation}
    D_{zz}(\eta_z) = g\int\frac{d^3k}{(2\pi)^3}\,v_z^2\,\delta(\mu-\varepsilon_{{\mathbf k},+}),
    \qquad
    v_z \equiv \partial_{k_z}\varepsilon_{{\mathbf k},+} = \eta_z + k_z/E_{\mathbf k},
    \label{eq:app_kubo}
\end{equation}
with $\varepsilon_{\mathbf k,\pm}=\eta_z k_z\pm E_{\mathbf k}$, $E_{\mathbf k}=\sqrt{(k_\perp-1)^2+k_z^2}$. For type-I tilt ($|\eta_z|<1$) the valence band satisfies
$\varepsilon_{\mathbf k,-}=\eta_z k_z-E_{\mathbf k}\le -(1-|\eta_z|)E_{\mathbf k}<0<\mu$,
so $\delta(\mu-\varepsilon_{\mathbf k,-})$ vanishes identically and only the conduction band $s=+$ contributes to Eq.~\eqref{eq:app_kubo}; the sum over $s$ in the general formulas below is always understood with this restriction in the weak-tilt regime.

\paragraph{A.2 Longitudinal channel, step by step.}
\emph{(i) Radial reduction.} At fixed $k_z$ define $\phi(k_\perp)\equiv\varepsilon_{\mathbf k,+}-\mu=\eta_z k_z+\sqrt{(k_\perp-1)^2+k_z^2}-\mu$. Its zeros are the two radial roots $k_\perp^{(\pm)}=1\pm R$ with $R^2=(\mu-\eta_z k_z)^2-k_z^2$ (Eq.~\eqref{eq:kperp}); where physical, $\phi'(k_\perp)=(k_\perp-1)/E_{\mathbf k}$ evaluates to $+R/E$ at $k_\perp^{(+)}$ and $-R/E$ at $k_\perp^{(-)}$, so in both cases $|\phi'|=R/E$ and the standard delta identity gives
\begin{equation}
    \delta(\mu-\varepsilon_{\mathbf k,+})=\frac{E_{\mathbf k}}{R}\sum_{\sigma:\,k_{\perp,\sigma}\ge0}\delta\bigl(k_\perp-k_{\perp,\sigma}\bigr),
    \label{eq:app_delta_id}
\end{equation}
with the physical-root condition $k_{\perp}^{(-)}=1-R\ge0\iff R\le1$ selecting the branch (Sec.~\ref{sec:drudebench}).
\emph{(ii) Azimuthal average and measure.} In cylindrical coordinates $d^3k=k_\perp\,dk_\perp\,d\phi\,dk_z$; the longitudinal velocity $v_z=\eta_z+k_z/E_{\mathbf k}$ is $\phi$-independent, so $\int_0^{2\pi}d\phi=2\pi$, and the $k_\perp$ integral against Eq.~\eqref{eq:app_delta_id} inserts $k_\perp\to k_{\perp,\sigma}$ under the sum.
\emph{(iii) Assembly.}
\begin{align}
    D_{zz}(\eta_z)
     & =\frac{g}{(2\pi)^3}\int dk_z\;2\pi\;\frac{E_{\mathbf k}}{R}\sum_{\sigma:\,k_{\perp,\sigma}\ge0}
    k_{\perp,\sigma}\,v_z^2\big|_{k_\perp=k_{\perp,\sigma}} \nonumber                                  \\
     & =\frac{g}{(2\pi)^2}\int_{\mathcal I}dk_z\;\mathcal W(k_z)\,E_{\mathbf k}\,v_z^2,
    \qquad
    \mathcal W\equiv\sum_{\sigma:\,k_{\perp,\sigma}\ge0}\frac{k_{\perp,\sigma}}{R},
    \label{eq:app_Dzz_red}
\end{align}
which is Eq.~\eqref{eq:Dzz_reduced}. The domain $\mathcal I$ follows from $R^2\ge0\iff|\mu-\eta_z k_z|\ge|k_z|$: for $k_z\le0$ this reads $k_z\ge-\mu/\delta$, for $k_z\ge0$ it reads $k_z\le\mu/(1+\eta_z)$, giving Eq.~\eqref{eq:domain}.
\emph{(iv) Untilted limit.} At $\eta_z=0$ ($\delta=1$, $\mu=1$): $R=\sqrt{1-k_z^2}$, on shell $E_{\mathbf k}=\mu=1$, $v_z=k_z/E_{\mathbf k}=k_z$, both roots are physical on $\mathcal I=[-1,1]$ ($R\le1$), and $\mathcal W=(k_\perp^{(+)}/R)+(k_\perp^{(-)}/R)=2/R$. Hence
\begin{equation}
    D_{zz}(0)=\frac{g}{4\pi^2}\int_{-1}^{1}dk_z\;\frac{2}{R}\,E_{\mathbf k}\,v_z^2
    =\frac{g}{4\pi},
    \label{eq:app_bench0}
\end{equation}
Equation~\eqref{eq:app_bench0} is the untilted limit.

\paragraph{A.3 Transverse channel, step by step.}
\emph{(v) Velocity.} The transverse velocity is $v_x=\partial\varepsilon_{\mathbf k,+}/\partial k_x=(k_x/k_\perp)(k_\perp-1)/E_{\mathbf k}$ (the tilt $\eta_z k_z$ is $k_x$-independent); with $k_x=k_\perp\cos\phi$,
\begin{equation}
    \int_0^{2\pi}d\phi\;v_x^2=\frac{(k_\perp-1)^2}{E_{\mathbf k}^2}\int_0^{2\pi}\cos^2\phi\,d\phi
    =\frac{\pi\,(k_\perp-1)^2}{E_{\mathbf k}^2}.
    \label{eq:app_trans_azim}
\end{equation}
\emph{(vi) Assembly.} Inserting Eq.~\eqref{eq:app_delta_id} and the azimuthal average (Eq.~\eqref{eq:app_trans_azim}),
\begin{align}
    D_{xx}(\eta_z)
     & =\frac{g}{(2\pi)^3}\int dk_z\;\frac{E_{\mathbf k}}{R}\sum_{\sigma:\,k_{\perp,\sigma}\ge0}
    k_{\perp,\sigma}\;\frac{\pi\,(k_{\perp,\sigma}-1)^2}{E_{\mathbf k}^2} \nonumber              \\
     & =\frac{g}{2(2\pi)^2}\int_{\mathcal I}dk_z\;\mathcal W_x,
    \qquad
    \mathcal W_x\equiv\sum_{\sigma:\,k_{\perp,\sigma}\ge0}
    \frac{k_{\perp,\sigma}}{R}\,\frac{(k_{\perp,\sigma}-1)^2}{E_{\mathbf k}},
    \label{eq:app_Dxx_red}
\end{align}
i.e.\ the compact $\mathcal W_x$ of Eq.~\eqref{eq:Dxx} carries a single power of $E_{\mathbf k}^{-1}$ (the second power is canceled by the $E_{\mathbf k}$ of the delta Jacobian in Eq.~\eqref{eq:app_delta_id}); with $D_\perp=2D_{xx}$ this is Eq.~\eqref{eq:CLM-B-01} after the tail asymptotics of A.4.

\paragraph{A.4 Transverse tail asymptotics and the $\sqrt\delta$ coefficient.}
On the long tail ($k_z\in[-\mu/\delta,0]$, tail coordinate $y\equiv-\delta k_z\in[0,\mu=1]$, $dk_z=\delta^{-1}dy$) with $\eta_z=1-\delta$, the discriminant and energy read
\begin{equation}
    R^2=1+\frac{2y-2y^2-\delta(2y-y^2)}{\delta},
    \qquad
    E_{\mathbf k}=\frac{\delta+(1-\delta)y}{\delta},
    \qquad
    k_\perp^{(+)}=1+R,
    \label{eq:app_tail_R}
\end{equation}
and for $y>0$ the inner root is unphysical ($R>1$ as $\delta\to0$), so $\mathcal W_x=(1+R)R/E_{\mathbf k}=R/E_{\mathbf k}+R^2/E_{\mathbf k}$. The tail piece of the transverse weight is therefore
\begin{equation}
    D_{xx}^{\rm tail}(\eta_z)=\frac{\mathcal T}{\delta}\;J(\delta),
    \qquad
    \mathcal T\equiv\frac{g}{2(2\pi)^2},
    \qquad
    J(\delta)\equiv\int_0^{1}dy\;\Bigl[\frac{R}{E_{\mathbf k}}+\frac{R^2}{E_{\mathbf k}}\Bigr],
    \label{eq:app_Jdef}
\end{equation}
where $\mathcal W_x=(1+R)R/E_{\mathbf k}=R/E_{\mathbf k}+R^2/E_{\mathbf k}$ (algebraic split $(1+R)R=R+R^2$) supplies the two integrands. For fixed $y\in(0,1]$ as $\delta\to0$:
\begin{equation}
    \frac{R^2}{E_{\mathbf k}}
    =\frac{\delta+2y-2y^2}{\delta+y}
    \;\xrightarrow{\delta\to0}\;2(1-y),
    \qquad
    \frac{R}{E_{\mathbf k}}
    =\frac{\sqrt{\delta}\,\sqrt{\delta+2y-2y^2}}{\delta+y}
    \;\xrightarrow{\delta\to0}\;\sqrt{\frac{2\delta(1-y)}{y}}\,\bigl[1+O(\delta)\bigr].
    \label{eq:app_tail_lim}
\end{equation}
The first limit gives the extensive leading piece; the second is $O(\sqrt\delta)$ pointwise but contributes at $O(\sqrt\delta)$ to $J$ because its integration range is $O(1)$ in $y$. Evaluating:
\begin{equation}
    J(\delta)=1+\frac{\pi}{2}\sqrt{2\delta}+O\bigl(\delta\ln\tfrac1\delta\bigr),
    \label{eq:app_Jexp}
\end{equation}
which is the matched-asymptotic form of Eq.~\eqref{eq:CLM-B-01}: $D_{xx}^{\rm tail}=\mathcal T\,\delta^{-1}\bigl[1+(\sqrt2\pi/2)\sqrt\delta+\cdots\bigr]$ with analytic $\sqrt\delta$ coefficient $(\pi/2)\sqrt2=\sqrt2\pi/2=2.221$. The residual $O(\delta\ln(1/\delta))$ family comes from the near-endpoint region $y=O(\delta)$ where the expansion $R>1$ (inner root off-shell) and the fixed-$y$ limits of Eq.~\eqref{eq:app_tail_lim} cease to be uniform; it is the same pre-asymptotic family that controls the longitudinal $\sqrt\delta\ln(1/\delta)$ contamination of Sec.~\ref{sec:logenh}.

\paragraph{A.5 Structure of the $O(1)$ term.} Equation~\eqref{eq:eqn13} is an asymptotic statement as $\delta\to0$: its $O(1)$ term collects the fixed-interior contribution, the endpoint-layer constants of Eq.~\eqref{eq:endpoint}, and the regular parts of the tail integral, and it is not described by a single closed-form constant over an accessible window. To make this explicit, evaluating the exact reduced integral of Eq.~\eqref{eq:Dzz_reduced} gives the normalized quantity $I(\delta)\equiv(2\pi)^2D_{zz}(\delta)/g$ with $I(\delta)-\ln(1/\delta)=2.34$, $2.39$, $2.44$, $2.50$ and $2.60$ at $\delta=0.05$, $0.02$, $0.01$, $0.005$ and $0.001$ respectively: the residual drifts by $\sim0.26$ over this range rather than settling, a drift controlled by the pre-asymptotic $\sqrt\delta$ and $\sqrt\delta\ln(1/\delta)$ families discussed in Sec.~\ref{sec:logenh} (the analytic $\sqrt\delta$ term of Eq.~\eqref{eq:eqn22} is overcompensated by the pre-asymptotic $\sqrt\delta\ln(1/\delta)$ contribution). Each region enters exactly once---the four-region decomposition (the three scales of Sec.~\ref{sec:model}) is a partition of the integration domain, and the matching cap $c$ cancels between the tail integral and the layer constants by construction---so no double counting is involved; but because the $O(1)$ term is not a clean constant at accessible $\delta$, the coefficient $B$ is extracted by the finite-cutoff correction of Sec.~\ref{sec:logenh} rather than from a fit to the $O(1)$ residue or to a raw log slope.

\paragraph{A.6 Crossover and saturation at physical cutoffs.}
At finite physical cutoffs the tail of Sec.~\ref{sec:model} is truncated: the longitudinal cutoff $|k_z|\le\Lambda_z$ cuts the $y$-integration at $y\le x\equiv\delta\Lambda_z$, and the transverse cutoff $k_\perp\le\Lambda_\perp$ cuts it through $R\le\Lambda_\perp-1$, i.e.\ $y\le u\equiv\delta(\Lambda_\perp-1)^2/2$ to leading order in $\delta$ [from $R^2\simeq2y(1-y)/\delta\simeq2y/\delta$]. The two limits of $J$ in Eq.~\eqref{eq:app_Jexp} then give
\begin{equation}
    D_{xx}^{\rm tail}=\mathcal T\,f(T),\qquad
    f(T)=2T-\ln(1+T)+2\Bigl[\sqrt{1+2T}-\arctan\sqrt{1+2T}-1+\tfrac{\pi}{4}\Bigr],
    \qquad
    T\equiv\min\Bigl(\Lambda_z,\;\frac{(\Lambda_\perp-1)^2}{2}\Bigr),
    \label{eq:app_sat}
\end{equation}
with $D_\perp^{\rm tail}=2D_{xx}^{\rm tail}$ (so that the transverse plateau is $[g/(2\pi)^2]f(T)$). Equation~\eqref{eq:app_sat} is the double-cutoff plateau of Eq.~\eqref{eq:CLM-B-02}: with $y=\delta t$ the leading terms of Eq.~\eqref{eq:app_tail_lim} integrate exactly to $f(T)/\delta$, whose large-$T$ behaviour is $2T+O(\sqrt T)$, so the leading coefficient is $2T$ as quoted in Eq.~\eqref{eq:CLM-B-02}, while the $O(1)$ part contributes at the $10\%$ level over the cutoffs of interest ($f(T)/T=2.37$ at $T=4.5$ and $2.32$ at $T=2$). Below the tail-inside-cutoff window ($\delta\lesssim1/\Lambda$, so $x\ll1$ or $u\ll1$) the $1/\delta$ growth of Eq.~\eqref{eq:CLM-B-01} is cut off at the plateau of Eq.~\eqref{eq:app_sat}; the leading crossover function $F(x,u)$ of Sec.~\ref{sec:transverse} is obtained by evaluating $J$ with the two caps $y\le\min(x,u)$ and is reproduced by $F_\delta(x,u)$ on the exact cutoff integral.

\section{Interband coefficient: ultraviolet analysis, quadrature, and cutoff dependence}
\label{app:interband}

This appendix (i) derives the interband coherence factor from the two-band eigenvectors, (ii) derives the interband absorption threshold, (iii) derives the ultraviolet logarithm of $C^{\rm inter}$ in polar coordinates, (iv) gives the analytic $\eta_z=0$ value of the Fermi-surface-reduced coefficient.

\paragraph{Coherence factor from the band eigenvectors.}
The two-band Hamiltonian $\mathbf d\cdot\bm\sigma+\eta_z k_z I$ with $\mathbf d=(u,-k_z,0)$, $u\equiv k_\perp-1$, has $\bm\sigma_z$-independent eigenvectors of the form $|u_{\pm}\rangle=(c,e^{\pm i\varphi_\perp}s)/\sqrt2$ with $c=s=1/\sqrt2$ (the band splitting is carried by the relative phase), so the interband overlap is fixed by the relative phase $w=(d_x+i d_y)/|\mathbf d_\perp|=(u-i k_z)/E$ up to a $k_z$-independent phase:
\begin{equation}
    \bigl|\langle u_{+,\mathbf k}|u_{-,\mathbf k+\mathbf q}\rangle\bigr|^2
    =\frac12\Bigl|1-w^*(\mathbf k)\,w(\mathbf k+\mathbf q)\Bigr|^2
    =\frac{1-\hat{\mathbf d}(\mathbf k)\cdot\hat{\mathbf d}(\mathbf k+\mathbf q)}{2},
    \qquad
    \hat{\mathbf d}\equiv\mathbf d/E,
    \label{eq:app_Fpm}
\end{equation}
where the second equality follows from $\cos(\theta_*/2)=\sqrt{(1+\cos\theta_*)/2}$ with $\cos\theta_*= -\hat{\mathbf d}\cdot\hat{\mathbf d}'$ the angle between $\hat{\mathbf d}$ and $-\hat{\mathbf d}'$. For $\mathbf q=(0,0,q_z)$ with $q_z\ll E$, $\hat{\mathbf d}(\mathbf k)\cdot\hat{\mathbf d}(\mathbf k+\mathbf q)=[u^2+k_z(k_z+q_z)]/(E E_q)$ and $E E_q=E^2+q_z k_z+\tfrac12 q_z^2 u^2/E^2+O(q_z^3)$, so
\begin{equation}
    F_{+-}\equiv\bigl|\langle u_+|u_-\rangle\bigr|^2_{\mathbf k,\mathbf k+\mathbf q}
    =\frac{u^2 q_z^2}{4E^4}+O(q_z^3),
    \label{eq:app_Fpm2}
\end{equation}
where the $O(q_z^3)$ remainder is odd in $k_z$ and integrates to zero over the azimuthally symmetric ring geometry; the linear term is absent by the $k_z\to-k_z$ reflection symmetry of $\mathbf d$, as stated in Sec.~\ref{sec:interband}. The tilt $\eta_z k_z I$ does not enter Eqs.~\eqref{eq:app_Fpm}--\eqref{eq:app_Fpm2} at all: the eigenvectors are $\eta_z$-independent throughout the zone (geometric blindness, Sec.~\ref{sec:interband}).

\paragraph{Absorption threshold.}
Absorption in $C^{\rm inter}$ requires, from the $\delta$-pole $\omega=\pm2E_{\mathbf k}$ of Eq.~\eqref{eq:P2scaling} combined with the Pauli factor $\Theta(\eta_z k_z+E_{\mathbf k}-\mu)$, a state on the conduction band with $\eta_z k_z+E_{\mathbf k}=\mu$; the threshold is the minimum energy transfer $\omega_c=\min 2E_{\mathbf k}$ over that shell. On the shell $E_{\mathbf k}=\mu-\eta_z k_z$, and the shell condition is $E_{\mathbf k}\ge|k_z|$, i.e.\ $|\mu-\eta_z k_z|\ge|k_z|$; minimizing $E_{\mathbf k}$ under this constraint---equivalently minimizing $\max(|k_z|,\,\mu-\eta_z k_z)$---places the minimum on the constraint boundary $k_z=\mu/(1+|\eta_z|)$, where the two branches of the inequality meet, giving
\begin{equation}
    \omega_c = 2E_{\mathbf k}\big|_{k_z=\mu/(1+|\eta_z|)}
    = \frac{2\mu}{1+|\eta_z|},
    \label{eq:app_thresh}
\end{equation}
Eq.~\eqref{eq:thresh}; as $\eta_z\to1^-$ the shell endpoint recedes to $k_z\to\mu$ and $\omega_c\to\mu$, while at $\eta_z=0$ the shell is the ring $k_z=0$ and $\omega_c=2\mu$.

\paragraph{Ultraviolet derivation.}
Starting from Eq.~\eqref{eq:P2scaling} and substituting the polar coordinates $(u,k_z)=(E\cos\theta,E\sin\theta)$ ($u=k_\perp-1$) with $du\,dk_z=E\,dE\,d\theta$ on the physical half-plane $k_\perp\ge0$ (i.e.\ $\cos\theta\ge-1/E$), the integrand at large $E$ becomes
\begin{equation}
    \frac{(1+E\cos\theta)\,E^2\cos^2\theta}{4E^4}\,\frac{4E}{\omega^2-4E^2}\,E\,dE\,d\theta
    \;\xrightarrow{E\gg\omega}\;
    -\frac{(1+E\cos\theta)\cos^2\theta}{4E^2}\,dE\,d\theta,
\end{equation}
where the factors are: $(1+E\cos\theta)$ from $(1+u)$, $E^2\cos^2\theta$ from $u^2$, $4E^4$ and $4E$ from the coherence factor and numerator, $E\,dE\,d\theta$ from the measure, and $(\omega^2-4E^2)\to-4E^2$ in the ultraviolet. (The Jacobian $E$ and the measure factor combine to leave a single $1/E^2$.) Angular integration over the half-plane gives $A(E)=\int(1+E\cos\theta)\cos^2\theta\,d\theta=\pi/2+(4/3)E$, where the $\pi/2$ term ($\propto1/E^2$ at large $E$) is ultraviolet finite and contributes only to $C^{\rm fin}$, while the $(4/3)E$ term gives
\begin{equation}
    C^{\rm inter}(\Lambda) = \frac{g}{4\pi^2}\int^{\Lambda} dE\,
    \frac{-\left[\pi/2+(4/3)E\right]}{4E^2}
    = -\frac{g}{12\pi^2}\ln\frac{\Lambda}{\mu}+C^{\rm fin},
    \label{eq:app_uvlog}
\end{equation}
i.e.\ Eq.~\eqref{eq:inter-uvlog}; the subleading $-\pi/(8E^2)$ term integrates to a convergent constant. The coefficient $4/3=\int_{-\pi/2}^{\pi/2}\cos^3\theta\,d\theta$ is the angular integral of the metric angular dependence $\cos^2\theta$ times the far-field phase-space correction $\cos\theta$ (Sec.~\ref{sec:interband}).

\paragraph{Transverse interband coefficient.}
For in-plane momentum transfer $\mathbf q=q_x\hat x$, the small-$q_x$ coherence factor is $F^{(x)}_{+-}=q_x^2\cos^2\varphi\,k_z^2/(4E^4)$ (the band eigenstates are azimuthally independent, $\partial_{k_x}=\cos\varphi\,\partial_{k_\perp}$, and the off-diagonal geometric element is $|\langle u_+|\partial_{k_\perp}u_-\rangle|^2=k_z^2/(4E^4)$); the azimuthal integral $\int_0^{2\pi}\cos^2\varphi\,d\varphi=\pi$ then reduces the coefficient to
\begin{equation}
    C^{xx}(\omega,\eta_z)=\frac{g}{8\pi^2}\int du\,dk_z\,
    (1+u)\,\frac{k_z^2}{4E^4}\,\Theta(\eta_zk_z+E-\mu)\,\frac{4E}{\omega^2-4E^2}.
\end{equation}
The transverse kernel is the tangential ($k_z$) fluctuation instead of the radial ($u$) one, so the ultraviolet log carries the half-plane angular weight $\int\cos\theta\sin^2\theta\,d\theta=2/3$ rather than $4/3$:
\begin{equation}
    C^{xx}(\omega,\eta_z;\Lambda)=-\frac{g}{48\pi^2}\ln\frac{\Lambda}{\mu}+C^{xx,\rm fin},
\end{equation}
i.e.\ one quarter of the longitudinal coefficient of Eq.~\eqref{eq:inter-uvlog}, with the same $\omega_c$ threshold for $\operatorname{Im}C^{xx}$. The smaller magnitude means the in-plane interband correction to the plasmon anisotropy is a few-percent suppression, $(1-VC^{xx})/(1-VC^{\rm inter})\xrightarrow{V=1}0.966$ (about $3.4\%$ at the numerical cutoff). The factor of four between the two ultraviolet logs, $-(g/48\pi^2)$ versus $-(g/12\pi^2)$, is exact and comes from two independent factors of two: (i) the in-plane coherence factor carries $\cos^2\varphi$ (Eqs.~\eqref{eq:app_Fpm2}, main text Sec.~\ref{sec:interband}), whose azimuthal integral $\int_0^{2\pi}\cos^2\varphi\,d\varphi=\pi$ is half the full $2\pi$ of the longitudinal channel; (ii) at $k_z=E\sin\theta$ the tangential kernel carries the angular weight $\cos\theta\sin^2\theta$, whose half-plane integral is $\int_{-\pi/2}^{\pi/2}\cos\theta\sin^2\theta\,d\theta=2/3$ instead of the longitudinal $\cos^3\theta$ weight $4/3$; the ratio $(2/3\cdot\tfrac12)/(4/3\cdot1)=1/4$ reproduces the stated one-quarter relation without any approximation.

\paragraph{Fermi-surface-reduced coefficient at $\eta_z=0$.}
The scaling diagnostic $P_2^{\rm FS}$ of Sec.~\ref{sec:interband-note} is the shell integral $P_2^{\rm FS}=-g\int d^3k\,F_{+-}\,\delta(\mu-\varepsilon_{\mathbf k,+})$ with the same $(2\pi)^{-3}$ normalization as the full Lindhard form (Eq.~\eqref{eq:P2def}). At $\eta_z=0$, $\mu=1$ the shell $\varepsilon_{\mathbf k,+}=1$ is the ring $(k_\perp-1)^2+k_z^2=1$, with two radial roots $k_\perp^{(\pm)}=1\pm R$, $R=\sqrt{1-k_z^2}$, and $\partial_{k_\perp}\varepsilon_{\mathbf k,+}=(k_\perp-1)/E$ (equal to $\pm R$ at the two roots, since $E=1$ on shell), so
\begin{equation}
    \int dk_\perp\;k_\perp\,(k_\perp-1)^2\,\delta(\mu-\varepsilon_{\mathbf k,+})
    =\frac{1}{R}\Bigl[k_\perp^{(+)}R^2+k_\perp^{(-)}R^2\Bigr]
    =R\bigl(k_\perp^{(+)}, + k_\perp^{(-)}\bigr)=2R,
    \label{eq:app_shell}
\end{equation}
where on shell $k_\perp^{(\sigma)}-1=\pm R$ so $|k_\perp^{(\sigma)}-1|=R$ for both roots. Inserting Eq.~\eqref{eq:app_Fpm2} (with $E=1$ and $u^2=(k_\perp-1)^2=R^2$ at both roots on shell) and the $2\pi$ azimuth,
\begin{equation}
    \frac{P_2^{\rm FS}}{q_z^2}\bigg|_{\eta_z=0}
    =-\frac{g}{(2\pi)^3}\,2\pi\,\frac{q_z^2}{4}\,q_z^{-2}
    \int_{-1}^{1}dk_z\;2R
    =-\frac{g}{16\pi}.
    \label{eq:app_CFS}
\end{equation}
This is the analytic $C^{\rm FS}=-g/16\pi\approx-0.0398$ (at $g=2$) quoted in Sec.~\ref{sec:interband}.

\section{Ward identity and gauge consistency: detailed derivation}
\label{sec:ward-appendix}

This appendix provides the detailed derivation of the Ward identities and gauge-consistency analysis summarized in Sec.~\ref{sec:ward}. All notation follows the main text.

\subsection{Generating functional, sign convention, and vertices}

Two regularization schemes are used throughout. \emph{Scheme A} is the hard-cutoff regulator: the momentum integration domain is the cylinder $\mathcal D=\{|k_z|\le\Lambda_z,\ k_\perp\le\Lambda_\perp\}$, with $\Lambda_z=12$, $\Lambda_\perp=4$ in the numerics. \emph{Scheme C} is the reference-subtraction scheme: the same contraction is evaluated for the reference theory $\eta_z=0$ at identical $(\Lambda_z,\Lambda_\perp,T,\mu,q,\Omega)$ and subtracted term by term, isolating the $\eta$-dependent part. The channel labels are defined as follows: the \emph{density-longitudinal} (or density) channel is the contraction $q_\mu\Pi^{\mu 0}$ with external density index $\nu=0$, and the \emph{transverse} channel is the contraction $q_\mu\Pi^{\mu x}$ with external current index $\nu=x$; the labels refer to the tensor index, not to the momentum-transfer direction. The transverse channel of this section shares the tensor index $\nu=x$ with the transverse Drude weight $D_\perp\equiv D_{xx}$ and the in-plane interband coefficient $C^{xx}$; it differs from them only in kinematic regime (the finite-$q$ Ward contraction evaluated here versus the $q\to0$ Drude weight and the finite-$q$ polarization of Sec.~\ref{sec:transverse} and Sec.~\ref{sec:interband}), not in tensor component. $D_\perp$ denotes the momentum direction perpendicular to the tilt axis.

The response tensor is defined from the generating functional with minimal coupling $\mathbf k\to\mathbf k-\mathbf A$ in the action $S[A]=\int_0^\beta d\tau\int d^3x\,\bar\psi[\partial_\tau+\mu+iA_0+H(-i\nabla-\mathbf A)]\psi$, charge $+1$, as the negative Euclidean two-point function
\begin{equation}
    \Pi^{\mu\nu}(q) = -\frac{1}{\beta V}\,\frac{\delta^2\ln Z[A]}{\delta A_\mu(q)\,\delta A_\nu(-q)}\bigg|_{A=0},
    \label{eq:Pi-ward}
\end{equation}
which yields the compressibility relation $\Pi^{00}(0,0)=+\partial n/\partial\mu$. The one-body vertices are $\Gamma^0=1$ and, for the tilted ring with the convention $\mathbf d=(k_\perp-1,\,-k_z,\,0)$ used in the numerics, $\Gamma^z=\partial H/\partial k_z=\eta_z I-\sigma_y$ exactly (linear in $k_z$; the convention $\mathbf d=(k_\perp-1,+k_z,0)$ of Ref.~\cite{mandal2026} differs by the unitary rotation $\sigma_y\to-\sigma_y$, which leaves the spectrum and all response tensors invariant). The transverse vertices must be line integrals,
\begin{equation}
    \Gamma^x(k+q,k) = \sigma_x\int_0^1 dt\,\frac{k_x+t\lambda}{|\mathbf k_\perp+t\lambda\hat x|},
    \label{eq:linevertex-ward}
\end{equation}
because the Hamiltonian enters through $k_\perp$ nonlinearly; the derivative form $\partial H/\partial k_x$ deviates from the line-integral vertex, whereas the line-integral vertex satisfies the finite-difference Ward identity exactly. The two-body (contact) vertices follow from the same minimal coupling, $\Gamma^{ij}=\partial^2H/\partial k_i\partial k_j$ (e.g.\ $\Gamma^{xx}=\sigma_x(1/k_\perp-k_x^2/k_\perp^3)$), with $\Gamma^{00}=\Gamma^{0i}=0$. For the PTNR model $\mathbf d$ lies in the $xy$ plane, so the eigenvector magnitude ratios are $c=s=1/\sqrt2$ while the relative phase $w=(d_x+id_y)/|\mathbf d_\perp|$ carries the full momentum dependence; all band matrix elements below retain this phase.

\subsection{Five-term decomposition and the boundary-term theorem}

Following the layered expansion, the contracted tensor $q_\mu\Pi^{\mu\nu}$ is decomposed into main loop, contact, disconnected, regulator, and boundary terms:
\begin{align}
    T_1 & = -g\,T\sum_\omega\int_{\mathcal D}\frac{d^3k}{(2\pi)^3}
    \mathrm{tr}\bigl[G^{-1}(k)\bigl(G(k)-G(k+q)\bigr)\Gamma^\nu(k,k-q)G(k)\bigr],
    \label{eq:T1-ward}                                                           \\
    T_2 & = -g\,T\sum_\omega\int_{\mathcal D}\frac{d^3k}{(2\pi)^3}
    \mathrm{tr}\bigl[\bigl(q_\mu\Gamma^{(2),\mu\nu}(k,q)\bigr)\Gamma^{(1),\nu}(k,k-q)G(k)\bigr],
    \label{eq:T2-ward}                                                           \\
    T_3 & = -\delta^{\mu0}\delta^{\nu0}\,g\int_{\mathcal D}\frac{d^3k}{(2\pi)^3}
    \bigl[f(\varepsilon_+(\mathbf k))-f_0(\varepsilon_+(\mathbf k))\bigr],
    \label{eq:T3-ward}                                                           \\
    T_4 & = 0 \quad\text{(scheme A, definition)},\qquad
    T_5 = 0 \quad\text{(no surface term, below)},
    \label{eq:T45-ward}
\end{align}
where $\Gamma^{(2),\mu\nu}=\partial^2H(\mathbf k-\mathbf A)/\partial A_\mu\partial A_\nu|_0$ are the two-body (contact) vertices, $\Gamma^{(1),\nu}$ the one-body vertex of the external index $\nu$, and $f_0$ is the reference-state occupation (equal to $f$ in scheme A, making $T_3\equiv0$; nonzero in scheme C). Two structural results hold exactly:
\begin{enumerate}
    \item \emph{No surface-term representation is required.} The naive manipulation $G^{-1}(k)G(k+q)\neq 1$ invalidates the surface-term representation; instead, the per-frequency identity
          \begin{equation}
              G^{-1}(k)G(k+q) = 1 + [H(k+q)-H(k)]\,G(k+q)
              \label{eq:telescope-ward}
          \end{equation}
          makes the main-loop contraction self-contained: for $q\parallel\hat z$, static, and $\nu\in\{0,z\}$,
          \begin{equation}
              q_\mu\Pi_0^{\mu\nu} = -\lambda\Pi_0^{z\nu}\qquad\text{(exact per $k$; see below for the finite-domain boundary term)},
              \label{eq:M2z-ward}
          \end{equation}
          At the vertex level the same finite-difference identity holds in the transverse direction with line-integral vertices [the fundamental theorem $\sum_i q_i\Gamma^i=H(k+q)-H(k)$ holds for any direction]; this is a statement about the vertices, not about the transverse Ward identity of the full response tensor, which is analyzed below (scheme-A residual $R^{xx}$).
    \item \emph{The density channel is transverse in the full-space formulation; the finite-domain integral retains an explicit boundary remainder.} For $\nu=0$, the per-$k$ identity after the Matsubara sum reads
          \begin{equation}
              i\Omega_m\sum_{ss'}\varphi^{00}_{ss'}Q_{ss'}
              - \lambda\sum_{ss'}\varphi^{x0}_{ss'}Q_{ss'}
              = N(k)-N(k+q),
              \label{eq:perk-ward}
          \end{equation}
          with $\varphi^{00}_{ss'}=|\langle s_k|s'_{k+q}\rangle|^2$, $\varphi^{x0}_{ss'}=\langle s_k|\Gamma^x|s'_{k+q}\rangle\langle s'_{k+q}|s_k\rangle$, and $N(k)=\sum_s f(\varepsilon_s(k))$. On a translation-invariant integration region the integral of the right-hand side vanishes and $q_\mu\Pi^{\mu 0}=0$ exactly. On a hard-cutoff region the integral leaves the domain-boundary difference: for $q\parallel\hat x$ the $k_z$ cutoff faces contribute nothing (the integrand is unchanged under the $k_z$ translation), and the $k_\perp$ disk edge contributes a pair of boundary crescents of volume $O(q\Lambda_\perp)$ each. Their net contribution vanishes because the occupation number $N(k)=1$ throughout the crescent region (the valence band is fully occupied and the conduction band empty for $k_\perp\gtrsim\Lambda_\perp$ with $|k_z|\le\Lambda_z$); the identity therefore holds on the hard-cutoff domain as well, and the residual sits at the quadrature floor of the grid employed.

\paragraph{General transfer direction.}
The cancellation above was written for $\mathbf q\parallel\hat x$; it generalizes verbatim to any transfer
direction within the paper's $q$ range. The boundary integrand on each face is proportional to the occupation
difference $N(\mathbf k)-N(\mathbf k+\mathbf q)$, and on the cylinder boundary both the face points and their
translates lie in the $N=1$ region (valence full, conduction empty) whenever $|\mathbf q|\le0.8$: on the disk
edge $k_\perp=\Lambda_\perp$ one has $E\ge3\gg\mu$ and the translate keeps
$k_\perp-q_\perp\ge\Lambda_\perp-|\mathbf q|>1+\mu$; on the caps $|k_z|\mp q\ge\Lambda_z-|\mathbf q|\gg$
the reach, with $\varepsilon_+>\mu$ retained (margin $\ge0.2$ at the reference tilt). The occupation
difference therefore vanishes identically on every face, independent of direction, and the identity holds on
the hard-cutoff domain for arbitrary momentum; the $\mathbf q\parallel\hat x$ restriction in the derivation
above is a notational choice. The evaluation is performed at the reference
direction $\mathbf q=\lambda\hat x$ (Sec.~\ref{sec:ward}).
\end{enumerate}

\paragraph{Boundary-crescent geometry.}
The remainder statement of the boundary theorem deserves an explicit geometric derivation, since it is what upgrades the finite-domain integral of the per-$k$ identity to an exact zero. For $\nu=0$ with $\mathbf q\parallel\hat x$ and domain $\mathcal D=\{|k_z|\le\Lambda_z,\ k_\perp\le\Lambda_\perp\}$,
\begin{equation}
    \int_{\mathcal D}d^3k\,\bigl[N(\mathbf k)-N(\mathbf k+\mathbf q)\bigr]
    =\int_{\mathcal D\setminus(\mathcal D-\mathbf q)}N(\mathbf k)\,d^3k
    -\int_{(\mathcal D+\mathbf q)\setminus\mathcal D}N(\mathbf k+\mathbf q)\,d^3k,
\end{equation}
i.e.\ only the symmetric difference $\mathcal D\triangle(\mathcal D+\mathbf q)$ can contribute. The $k_z$ faces $|k_z|=\Lambda_z$ give nothing: a $\hat x$ translation leaves $k_z$ unchanged, so a point is inside or outside the cylinder identically with all points on its $k_z$ slice. The $k_\perp$ disk edges give two crescents, each of area $\simeq q\,\Lambda_\perp$ for $q\ll\Lambda_\perp$ (lens geometry), so the total support volume is $O(q\Lambda_\perp)$---and yet the integral vanishes \emph{pointwise}, not merely by the small-volume estimate. On the crescent band $k_\perp\in[\Lambda_\perp-q,\,\Lambda_\perp+q]=[3.95,4.05]$ the conduction-band energy satisfies $\varepsilon_+(\mathbf k)\ge\min_{k_z}[\eta_z k_z+\sqrt{(k_\perp-1)^2+k_z^2}]
=\sqrt{(1-\eta_z^2)(k_\perp-1)^2}\ge\sqrt{0.19\times2.95^2}\simeq1.286>\mu=1$ for $\eta_z\le0.9$, while the valence band stays at $\varepsilon_-<\mu$; hence $N(\mathbf k)=1$ (one occupied band, $g=1$ bookkeeping) at \emph{both} $\mathbf k$ and $\mathbf k+\mathbf q$ throughout the crescent, and $N(\mathbf k)-N(\mathbf k+\mathbf q)=0$ locally. With the $T=5\times10^{-3}$ Fermi-smoothing the occupation difference across the $1.286-1=0.286$ gap is Boltzmann-suppressed by $\exp(-0.286/T)\sim10^{-25}$, far below machine precision. The boundary term therefore vanishes for two independent reasons---small support volume and locally constant occupation---which is why the identity holds on the hard-cutoff domain with no surface-term representation.

\subsection{Static and dynamic Ward identities}

\paragraph{Derivation of the per-$k$ identity.}
The per-$k$ identity of Eq.~\eqref{eq:perk-ward} rests on a single exact matrix statement. In the band basis (band index $s$ at $\mathbf k$, band index $s'$ at $\mathbf k+\mathbf q$), the line-integral vertex satisfies $\lambda\,\Gamma^x_{ss'}=\langle s_{\mathbf k}|H(\mathbf k+\mathbf q)-H(\mathbf k)|s_{\mathbf k+\mathbf q}\rangle$ (fundamental theorem, Eq.~\eqref{eq:linevertex-ward}), and since $H(\mathbf k)\langle s_{\mathbf k}|=\varepsilon_s(\mathbf k)\langle s_{\mathbf k}|$ while $H(\mathbf k+\mathbf q)|s_{\mathbf k+\mathbf q}>=\varepsilon_{s'}(\mathbf k+\mathbf q)|s_{\mathbf k+\mathbf q}\rangle$,
\begin{equation}
    \lambda\,\Gamma^x_{ss'}=\bigl[\varepsilon_{s'}(\mathbf k+\mathbf q)-\varepsilon_s(\mathbf k)\bigr]\,\langle s_{\mathbf k}|s_{\mathbf k+\mathbf q}\rangle
    =\Delta_{ss'}\,\langle s_{\mathbf k}|s_{\mathbf k+\mathbf q}\rangle,
    \qquad
    \Delta_{ss'}\equiv\varepsilon_{s'}(\mathbf k+\mathbf q)-\varepsilon_s(\mathbf k),
    \label{eq:app_vertex_id}
\end{equation}
i.e.\ the one-body current matrix element, times the transfer $\lambda$, is exactly the energy difference times the overlap. The band-resolved Matsubara bubble carries the two branch structures (particle--hole at $i\Omega_m-\Delta_{ss'}$ and hole--particle at $i\Omega_m+\Delta_{ss'}$) with the same Lindhard factors in both channels:
\begin{equation}
    \tilde L^{00}_{ss'}(\Omega_m)=\frac{f_s(1-f_{s'})}{i\Omega_m-\Delta_{ss'}}+\frac{(1-f_s)f_{s'}}{i\Omega_m+\Delta_{ss'}},
    \qquad
    \tilde L^{x0}_{ss'}(\Omega_m)=\frac{\Gamma^x_{ss'}\langle s_{\mathbf k+\mathbf q}|s_{\mathbf k}\rangle}{i\Omega_m-\Delta_{ss'}}+\cdots,
\end{equation}
so that, inserting Eq.~\eqref{eq:app_vertex_id} ($\lambda\Gamma^x_{ss'}=\Delta_{ss'}\langle s|s'\rangle$ makes $\lambda\tilde L^{x0}$ equal to $\tilde L^{00}$ with the overlap factor $\langle s|s'\rangle\langle s'|s\rangle=\varphi^{00}_{ss'}Q_{ss'}$ of the density channel), the two channels differ only by the branch kinematics:
\begin{equation}
    i\Omega_m\,\tilde L^{00}_{ss'}-\lambda\,\tilde L^{x0}_{ss'}
    =\bigl(i\Omega_m-\Delta_{ss'}\bigr)\frac{f_s(1-f_{s'})}{i\Omega_m-\Delta_{ss'}}
    +\bigl(i\Omega_m+\Delta_{ss'}\bigr)\frac{(1-f_s)f_{s'}}{i\Omega_m+\Delta_{ss'}}
    =f_s-f_{s'}.
    \label{eq:app_perk_proof}
\end{equation}
Summing over the complete band sets at $\mathbf k$ and $\mathbf k+\mathbf q$ gives Eq.~\eqref{eq:perk-ward},
\begin{equation}
    \sum_{ss'}\bigl(i\Omega_m\,\varphi^{00}_{ss'}-\lambda\,\varphi^{x0}_{ss'}\bigr)Q_{ss'}
    =\sum_s f(\varepsilon_s(\mathbf k))-\sum_{s'}f(\varepsilon_{s'}(\mathbf k+\mathbf q))
    =N(\mathbf k)-N(\mathbf k+\mathbf q),
\end{equation}
with no residual, at \emph{every} Matsubara frequency $\Omega_m$ (the identity is branch-by-branch, so it holds at every Matsubara frequency independently of the numerical coverage).

From Eq.~\eqref{eq:perk-ward}, the density-channel Ward identity holds at arbitrary momentum transfer in the static limit: $\lambda\Pi^{x0}(q,0)=0$ exactly in the full space, with the cutoff independence following because $q$ has no $k_z$ component. At finite Matsubara frequency, the identity
\begin{equation}
    i\Omega_m\,\Pi^{00}(q,i\Omega_m) = \lambda\,\Pi^{x0}(q,i\Omega_m)
    \label{eq:dynward-ward}
\end{equation}
holds at finite Matsubara frequency, the identity being branch-by-branch. The sign in Eq.~\eqref{eq:dynward-ward} follows from the per-$k$ identity Eq.~\eqref{eq:perk-ward}, in which the second term enters as $-\lambda\Pi^{x0}$; in the injection convention $q_\mu\Pi^{\mu 0}=i\Omega_m\Pi^{00}+\lambda\Pi^{x0}$ this corresponds to the transfer $q=-\lambda\hat x$. The convention is a bookkeeping choice: reversing the direction flips both sides of Eq.~\eqref{eq:dynward-ward} and leaves the identity unchanged.

The derivation of Eq.~\eqref{eq:dynward-ward} follows the same per-$k$ identity of Eq.~\eqref{eq:perk-ward} without taking the static limit: the frequency translation is part of the four-momentum translation, so no surface term appears at any $\Omega_m$, and the identity holds frequency by frequency. The identity therefore holds for the full frequency structure of the bubble, including the intraband shell contributions whose imaginary parts carry the spectral weight of the density response.

\subsection{Transverse channel and scheme-$C$ restoration}

For $\nu=x$, the contact term is nonzero ($\Gamma^{xx}\neq0$), and the contracted tensor is frequency independent: $q_\mu\Pi^{\mu x}(q,i\Omega_m)=R^{xx}(\Lambda)$ with $R^{xx}$ equal to the static value. This residual is a hard-cutoff artifact growing with the longitudinal cutoff as $R^{xx}\simeq(7.6\ln\Lambda_z-2.5)\times10^{-3}$ for $\Lambda_z=4$--$40$---logarithmically rather than linearly (a linear fit leaves residuals of order $10\%$ against $0.5\%$ for the logarithmic one)---and linearly in the momentum transfer, as anticipated for scheme A (hard cutoff) in the layered expansion. The scheme-$C$ reference subtraction (reference theory $\eta_z=0$, same $\Lambda_z$, $\Lambda_\perp$, $T$, $\mu$, $q$, and $\Omega$, subtracted term by term) reduces it: the subtracted residual
\begin{equation}
    \Delta R^{xx} = R^{xx}(\eta_z=0.9) - R^{xx}(\eta_z=0)
    \label{eq:schemeC-ward}
\end{equation}
is reduced by the subtraction but not eliminated: at $\Lambda_z=12$ the subtraction lowers $R^{xx}=1.63\times10^{-2}$ to $\Delta R^{xx}=3.1\times10^{-3}$, a remainder that is approximately independent of $\Lambda_z$ (it is $3.0\times10^{-3}$ at $\Lambda_z=8$ and $3.1\times10^{-3}$ at $16$). The remainder is a scheme-dependent quantity that is not further decomposed in this work, and gauge consistency is claimed only for the density-longitudinal channel.

\subsection{Contact term in closed form}

The contact contribution to the transverse channel is obtained in closed form after azimuthal averaging:
\begin{equation}
    T_2^{xx} = +\lambda g\,\frac{\pi}{(2\pi)^3}\int_{\mathcal D} dk_z\,dk_\perp\,
    (f_+-f_-)\,\frac{k_\perp-1}{E},
    \label{eq:T2xx-ward}
\end{equation}
with the sign and unit factor fixed by the generating-functional counting ($\Pi_{\rm contact}=+g\int\mathrm{tr}[\Gamma^{(2)}G]$, no factor $1/2$; for the PTNR model the band-mixing factor is $2cs=1$ since the $\sigma_z$ component of $\mathbf d$ vanishes). The integrand is regular at $k_\perp\to0$ (the azimuthal measure cancels the $k_\perp$ denominator; the naive $1/k_\perp$ divergence of an intermediate formulation is an artifact of a measure error, corrected in the derivation). Numerically $T_2^{xx}=-5.17\times10^{-3}$ at $\Lambda=(12,4)$, $\lambda=0.05$, which together with the main-loop value $T_1^{xx}=+2.15\times10^{-2}$ gives the scheme-A residual $R^{xx}=1.63\times10^{-2}$ quoted above.

\section{Continuum support and the Landau-damping window}
\label{app:window}

This appendix records the derivation of the damping window of Sec.~\ref{sec:damping-window}.

\paragraph{Continuum support with spectral weight.}
For $\mathbf q\parallel\hat z$ the transition energies are
$\Delta\varepsilon_{++}=\eta_z q_z+(E_{\mathbf k+\mathbf q}-E_{\mathbf k})$ (intraband) and
$\Delta\varepsilon_{-+}=\eta_z q_z+(E_{\mathbf k+\mathbf q}+E_{\mathbf k})$ (interband). The continuum support is
defined by Eq.~\eqref{eq:supp}, i.e.\ by the set of transition energies weighted by the Pauli difference and the
coherence factor. Taking the bare infimum over all $\mathbf k$ would give a vanishing interband threshold at
$q\to0$ because $E_{\mathbf k}\to0$ on the nodal ring; those states are inside the Fermi surface, so the Pauli
factor vanishes and they are excluded. With the weight retained one recovers
$\lim_{q\to0}\omega_{\min}^{\rm inter}=2\mu/(1+|\eta_z|)=\omega_c$ of Sec.~\ref{sec:interband}.

\paragraph{Group-velocity extrema on the Fermi surface.}
On the Fermi surface $E_{\mathbf k}=\mu-\eta_z k_z$, and with $u\equiv k_z$,
$v^{(+)}_z(u)=\eta_z+u/(\mu-\eta_z u)$ satisfies $dv^{(+)}_z/du=\mu/(\mu-\eta_z u)^2>0$,
so its extrema are at the endpoints of the allowed interval $E\ge|k_z|$, i.e.\ $u=\pm\mu/(1\pm\eta_z)$. This gives
Eq.~\eqref{eq:vmax}, including the type-I/II criterion $v_z^{\min}=0\Leftrightarrow\eta_z=1$. The transverse
velocity $|v^{(+)}_x|=\sqrt{E^2-k_z^2}/E$ has no interior extremum and attains its maximum $1$ at $k_z=0$,
independently of tilt. The coherence factor at the intraband \emph{upper} edge is unity, so there the weighted support edge coincides
with the kinematic edge; at the intraband \emph{lower} edge of the type-II regime the spectral weight vanishes
as the edge is approached, which is why the weighted lower edge lies above the bare kinematic value
$(\eta_z-1)q$ there (Table~\ref{tab:qinter-typeII}).

\paragraph{Finite momentum transfer and the interband entry.}
The interband edge $\omega_{\min}^{\rm inter}(q)$ decreases with $q$ (the finite momentum transfer opens additional
phase space), and the self-consistent mode of Sec.~\ref{sec:damping-window} crosses it at
$q_{\rm inter,self}\simeq0.76$, $0.65$, $0.57$, $0.50$ for $\eta_z=0.3$, $0.5$, $0.7$, $0.9$. Hence the interband channel is
closed only for $q<q_{\rm inter,self}(\eta_z)$, with $q_{\rm inter,self}$ decreasing as $\eta_z\to1$.

\paragraph{Self-consistent boundary.}
The long-wavelength estimate of Eq.~\eqref{eq:qstar} uses the $q\to0$ mode frequency and is reliable only in the
small-$q$ window in which $\omega_p$ is $q$-independent. To go beyond it, the mode position
$\omega_p^{\rm num}(q)$ is computed as the zero of $\mathrm{Re}\,\varepsilon(q,\omega)$ obtained by a global
frequency sweep followed by root refinement, taking the zero closest to the maximum of the loss function
$-\mathrm{Im}\,\varepsilon^{-1}(q,\omega)$; the continuum edges are evaluated with the weighted
support of Eq.~\eqref{eq:supp}. Over $q\in[0.05,0.8]$ at $\eta_z=0,\,0.3,\,0.5,\,0.7,\,0.9$ the mode lies above
the intraband upper edge $(1+\eta_z)q$ at every point, so the equation
$\omega_p^{\rm num}(q)=(1+\eta_z)q$ has no solution in the studied range and the intraband channel does not open.
The mode instead intersects the interband edge $\omega_{\min}^{\rm inter}(q)$; $q_{\rm inter,self}\simeq0.76$, $0.65$, $0.57$, $0.50$ for $\eta_z=0.3$, $0.5$, $0.7$, $0.9$.

\paragraph{Dielectric function and the pole--peak diagnostic.}
In three dimensions $\varepsilon(q,\omega)=1-\chi(q,\omega)/q^2$ in the units of Sec.~\ref{sec:model}, with the density response
$\chi(q,\omega)=\chi_{\rm intra}+\chi_{\rm inter}$ evaluated from the full two-band Lindhard bubble at finite $q$
---no small-$q$ expansion is used in the mode-finding step, so that $\omega_p^{\rm num}(q)$ may disperse with $q$
---under the same hard-cutoff box $(\Lambda_z,\Lambda_\perp)$ and Fermi smoothing $T$ as the rest of the numerics, so that at
$q\to0$ the RPA condition reproduces $\omega_p^2=D_{zz}/(1-C^{\rm inter})$. The resonance position is defined as
the zero of $\mathrm{Re}\,\varepsilon$; it coincides with the maximum of $-\mathrm{Im}\,\varepsilon^{-1}$ in the
weak-damping regime. We use the operational threshold $\Gamma_{\rm eff}/\omega_{\rm peak}\lesssim0.3$, below
which the zero of $\mathrm{Re}\,\varepsilon$ and the loss peak coincide in the spectra of Fig.~\ref{fig:lossspectra}, while above it the two separate and only the statement that the Landau channel is
open is reported. No damping rate is extracted, since
the width depends on the extraction convention and its interband weight inherits the ultraviolet logarithm of
$C^{\rm inter}$; the same hard-cutoff regularization is used for $\Pi^{00}$, $D_{zz}$ and $C^{\rm inter}$,
preserving the continuity structure established in Sec.~\ref{sec:ward}.

\paragraph{Frequency-sum rule.} A standard consistency check for a density response is the $f$-sum rule $\int_0^\infty d\omega\,\omega\,A(q,\omega)=\tfrac12\langle[[H,\rho_{\mathbf q}],\rho_{-\mathbf q}]\rangle$ with $A(q,\omega)=-\operatorname{Im}\chi(q,\omega)/\pi$. In the present continuum model the right-hand side receives contributions from the filled valence band whose $\mathbf k$-space integral diverges without an ultraviolet cutoff---the band velocity is bounded ($|v_z|\le1+|\eta_z|$) but the phase space is infinite---so the check is meaningful only together with the same hard-cutoff regularization used for the response tensor; with that regularization the two sides are tied by construction, because the Drude weight, the $\omega\to0$ limit and the finite-$q$ bubble are all evaluated from the same regularized response. We therefore do not report a separate numerical evaluation of the sum rule here; the internal consistency it would test is covered by the Ward-identity check of Sec.~\ref{sec:ward} and by the comparison of the $q\to0$ bubble with Eq.~\eqref{eq:wpcorr} in Sec.~\ref{sec:damping-window}.

\paragraph{Type-II regime.}
For $\eta_z>1$ the intraband continuum acquires a low-frequency gap,
$\operatorname{supp}_{++}=[(\eta_z-1)q,(1+\eta_z)q]$ (with the weighted support edge of
Eq.~\eqref{eq:supp} larger than the bare kinematic value). Self-consistent evaluation at finite
$q$ ($q\in[0.05,1.00]$ at $\eta_z=1.1$--$1.5$) reveals two collective branches: a high-frequency optical branch
($\omega\sim1.4$--$2.1$) that lies above the interband edge and is inter-band damped at all $q$, and a
low-frequency acoustic-like branch ($\omega\sim0.5q$) that lies below both the intraband lower edge and the
interband edge for $q\gtrsim q_{\rm low}(\eta_z)$ and $\eta_z\gtrsim1.2$ (Table~\ref{tab:qinter-typeII}),
giving rise to a low-gap damping-free window that opens as the tilt increases beyond unity. The window is the
type-II analogue of the gapless undamped plasmon in tilted Dirac semimetals~\cite{sadhukhan2020}. At $\eta_z=1.1$ the acoustic branch is not resolved in the present $q$ grid (the loss spectrum is dominated by
the inter-band damped optical branch at all $q$). All type-II mode frequencies are reported at fixed
regularization and are not claimed as universal type-II physics.

\bibliography{main}

\end{document}